\documentclass[draftcls,11pt,onecolumn,english,twoside]{IEEEtran}

\usepackage{amssymb}
\usepackage[cmex10]{amsmath}
\usepackage{mathtools}
\usepackage[utf8]{inputenc}
\usepackage{graphicx}
\usepackage{caption}
\usepackage{subcaption}

\usepackage{cite}
\usepackage{url}

\newcommand{\rrn}{r_{r}}
\newcommand{\receiver}{\Omega_{r}}

\newcommand{\cdfNhit}{F(\receiver, t|r_0)}

\newcommand{\pdfNhit}{h(\receiver, t|r_0)}
\newcommand{\pdfNhitt}[1]{h(\receiver, #1|r_0)}
\newcommand{\dortPiDe}{\sqrt{4\pi D t^3}}
\newcommand{\expoPiDe}{\,e^{-d^2/4Dt}}

\newcommand{\pdfDecompose}{h(\receiver, \lambda, t|r_0)}
\newcommand{\pdfDecomposeAtT}[1]{h(\receiver, \lambda, #1 |r_0)}
\newcommand{\hitBeforeDecompose}{F_{}(\receiver, \lambda |r_0)}
\newcommand{\hitBeforeDecomposeT}{F_{}(\receiver, \lambda, t |r_0)}
\newcommand{\hitBeforeDecomposeAtT}[1]{F_{}(\receiver, \lambda, #1 |r_0)}

\newcommand{\channelResponse}[2]{F_c(\receiver, \lambda, #1, #2 |r_0)}

\newcommand{\channelResponseSumFromTo}[2]{F_{s_{#1:#2}}}
\newcommand{\rateOfISI}[1]{\text{ITR}}

\newcommand{\tsym}{t_s}

\newcommand{\symi}[1]{s_{#1}}
\newcommand{\symsq}[2]{s_{#1:#2}}
\newcommand{\decodedsi}[1]{\hat{s_{#1}}}
\newcommand{\decodedSRV}{\hat{S}}

\newcommand{\ntxs}{N^{\text{Tx}}}
\newcommand{\ntxsi}[1]{N^{\text{Tx}}_{#1}}
\newcommand{\nrxts}[2]{N^{\text{Rx}}(#1,#2)}
\newcommand{\Yts}[1]{N^{\text{Rx}}_{#1}}
\newcommand{\Ytsprev}[2]{N^{\text{Rx}}_{#1,#2}}

\newcommand{\pulsept}{t_{\text{peak}}}
\newcommand{\pulsepn}{n_{\text{peak}}}
\newcommand{\halfLife}{\Lambda_{1/2}}
\newcommand{\halfLifeT}[1]{\Lambda_{1/2} = #1 ~s}

\newcommand{\tpeakdegradation}{\frac{\sqrt{36D^2 + 16 D d^2 \lambda} - 6D}{8D \lambda}}

\newcommand{\perr}{Pe}
\newcommand{\pe}[1][]{Pe_{|#1}}
\newcommand{\pc}[1][]{Pc_{|#1}}
\newcommand{\ith}{i^{th}}

\newcommand{\mth}{m^{th}}

\newcommand{\etal}{\text{\textit{et al.}}}

\graphicspath{{../graphs}}

\title{Effect of Degradation in Molecular Communication: Impairment or Enhancement?}

\author{Akif Cem Heren, H. Birkan Yilmaz, \\Chan-Byoung Chae,~\IEEEmembership{Senior Member,~IEEE}, and Tuna Tugcu,~\IEEEmembership{Member,~IEEE}
\thanks{A. C. Heren and T. Tugcu are with NETLAB, Department of Computer Engineering, Bogazici University, Istanbul, 34342, Turkey (e-mail: akif.heren@boun.edu.tr; tugcu@boun.edu.tr).}% <-this % stops a space
\thanks{H. B. Yilmaz and C.-B. Chae are with the School of Integrated Technology, Yonsei Institute of Convergence Technology, Yonsei University, Korea (e-mail:
birkan.yilmaz@yonsei.ac.kr; cbchae@yonsei.ac.kr).}% <-this % stops a space
}

\begin{document}
\maketitle

\begin{abstract}
In the nanonetworking literature, many solutions have been suggested to enable the nanomachine-to-nanomachine communication. Among these solutions, we focus on what constitutes the basis for molecular communication paradigms --molecular communication via diffusion (MCvD). In this paper, we start with an analytical modeling of a spherical absorbing receiver under messenger molecule degradation and show that our formulations are in agreement with the simulation results of a similar topology. Next, we identify how such signal characteristics as pulse peak time and pulse amplitude are affected by degradation. Indeed, we show analytically how in MCvD, signal shaping is achieved through degradation. We also compare communication under messenger molecule degradation with the case of no-degradation and electromagnetic communication in terms of channel characteristics. Lastly, we evaluate the communication performance of the scenarios having various degradation rates. Here, we assess the system performance according to traditional network metrics such as the level of inter-symbol interference, detection performance, bit error rate, and channel capacity. Our results indicate that introducing degradation significantly improves the system performance when the rate of degradation is appropriately selected. We make a thorough analysis of the communication scenario by taking into account different detection thresholds, symbol durations, and communication distances.
\end{abstract}


\section{Introduction}

%Nano-communication has emerged in response to impending need of communicating between nano-machines for 

As nano-technology started to prevail, a need for communication between nano-scale machines has emerged. The growing field of nano-communication envisions a world where interaction between such machines is enabled, through which realization of complex tasks with macro-scale results is achieved. Molecular communication is a promising candidate to achieve this target, with approaches including molecular communication via diffusion (MCvD), calcium signaling, micro-tubules, pheromone and bacterial signaling~\cite{nakano2013molecularC}. Among these, MCvD has shown to be both an effective and energy-efficient strategy in the literature \cite{kuran2010energy, hiyama2006molecular, kim2013novelMT}. 

An MCvD system is composed of modulation, emission (transmission), signal propagation, reception and demodulation processes. In MCvD, the transmitters reside in a fluid medium and they emit molecules that modulate information. In the literature, information is demonstrated to be modulated on various aspects of the messenger molecules, such as molecule identity, molecule concentration, and signal frequency \cite{kuran2011modulation, srinivas2012molecular,yilmaz2014simulationSO}. The information is transmitted via the propagation (diffusion) of emitted molecules through the environment. Diffusion causes the molecules to propagate and spread throughout the environment. The propagation is generally considered to be restricted to diffusion, unless the environment has flow currents as in \cite{srinivas2012molecular, parcerisa2009molecular}. Finally, at the receiver side, the molecules react with the receptors over the receiver node surface and are removed from the environment. A macro-scale counterpart of such a system was introduced by Farsad $\etal$ in \cite{farsad2014molecularCL}. In the nature, most of the receptor types remove the information-carrying molecules from the environment once they arrive at the receiver~\cite{cuatrecasas1974membraneR, kuran2013tunnel}. Therefore, in most of the cases, almost all molecules contribute to the signal once. If the receptors do not remove molecules from the environment, receiving cells have other mechanisms to guarantee that each molecule contributes to the signal only once (e.g., acetylcholinesterase (ACh) breaks down the molecules in the neuromuscular junctions)~\cite{kuran2013tunnel}. For the reception process, the literature has mostly considered an absorbing receiver in a 1-D environment \cite{nakano2012channel} or a  hypothetical sphere in a 3-D environment with the ability to measure concentration within \cite{pierobon2013capacity}, until the authors of \cite{ThreeDimForAbsorbingReceiver} formulated channel response for 3-D molecular communications with an absorbing spherical receiver.

In the nature, the most typical use of molecular degradation is found in the synaptic cleft between axons and dendrites of the neurons in the brain. This communication is needed to carry the action potential from the axon of the pre-synaptic neuron to the dendrite of the post-synaptic neuron~\cite{mesiti2013nanomachineTN, liu2014interSI}. In this process, the pre-synaptic neuron, following an action potential reaching to the synapse, releases bursts of ACh molecules to the synaptic cleft. These ACh molecules diffuse through the synaptic junction and result in a new action potential in the post-synaptic neuron after absorption.\footnote{For a single burst, approximately $10^{-17} $ moles of ACh is released} The degradation in this case is introduced via the ACh-hydrolyzing enzyme Acetylcholinesterase (AChE), which terminates the ACh action in the synaptic junctions. A single AChE molecule can terminate $6 \times 10^{5}$ ACh molecules per minute \cite{goodman2006goodman}. This is one of the fastest cases of degradation in the nature and it is close to the theoretical limits of enzymatic hydrolysis \cite{tai2003finite}.

By cleaning the communication channel from the messenger molecules, AChE enables the communication in the synaptic cleft to continue as new waves of messenger molecules arrive in the channel and allows consecutive triggering of the contraction of the muscles such as for the beating of the heart. From a networking point of view, the presence of AChE reduces inter symbol interference (ISI) and greatly increases overall bit rate of the system. Without the AChE presence, the channel would overflow with ACh, stopping all the communication in the synaptic cleft. Therefore, the degradation caused by AChE is crucial to human life. Historically, anti-AChE agents (a.k.a nerve gases) have been used to inhibit AChE activity as a deadly weapon \cite{bignami1975behavioral}. 

In the nanonetworking literature, molecular degradation has been considered conservatively. Although evidence in the nature can be found for cases where degradation increases communication performance, the literature is divided when it comes to considering degradation as detrimental or beneficial. 

In \cite{nakano2012channel}, Nakano $\etal$ investigated the channel capacity of communication via diffusion, considering the exponential degradation of the messenger molecules. Although they provide a solid formulation for the channel capacity, they only analyzed a limited amount of scenarios with few performance metrics. In addition, their channel model was a 1-D approximation, failing to comprehensively model the effects of degradation in the system. In \cite{arifler2011capacity}, a similar analysis was made for the case of a degradation pattern following a Weibull distribution, again using the 1-D absorber approximation. 

In \cite{liu2014channel}, Liu $\etal$ provided a channel capacity analysis for the molecular communication via diffusion with a receiver with ligand receptors. The use of ligand receptors entitles the consideration of binding and release rates of the messenger molecules to the receptors and the exponential degradation of these molecules in the propagation medium. In this study, the benefits of degradation is overlooked by considering only a very limited set of cases. Additionally, this model uses a concentration-based receiver approximation for the channel response formulation. 

In \cite{wang2014transmitPS}, Wang $\etal$ introduced secondary molecules to cancel the effect of the primary molecules i.e., to shape the transmit signal. The first hitting formulation of a 1-D environment is, however, used and the process of physical cancellation is not sufficiently studied.

In, \cite{noel2013improving, noel2013using, noel2014improvingRPJournal}, Noel $\etal$ provided a thorough analysis of the effects of enzymatic degradation of messenger molecules by modeling enzymatic reactions according to Michaelis-Menten mechanism, which is a highly accepted model of enzymatic degradation. Although this model covers the basis of benefits of degradation by stating the signal shaping aspect, it does not deliver a complete mathematical derivation regarding the reason behind the improvements. In the study, the analysis is based on a single symbol duration and does not show the benefits of degradation in an ongoing communication scenario. Finally, this study utilizes an enhanced version of concentration-based receiver, where the receiver is no longer a point in space but a finite volume (a sphere or a rectangular prism), which counts the number of molecules in bulk. In this model, the receiver node does not absorb or manipulate the messenger molecules. 

In this paper, we derive the channel response function considering exponential degradation in a 3-D environment for the case of an absorbing receiver. To the best of our knowledge, \emph{this is the first analytical consideration of molecular degradation in a 3-D environment with an absorbing spherical receiver}. Most prior work in the  literature only considers 1-D environments, where limited investigation is done regarding the effects of degradation. In this work, we derive a solid analytical formulation for the fraction of received molecules in the case of a 3-D absorbing receiver with messenger molecule degradation. Moreover, we provide peak time and peak amplitude formulations for the received signal to analytically investigate the effect of degradation on channel characteristics. Further, we elaborate on the effects of degradation on system performance using traditional network metrics; receiver operating characteristic (ROC), bit error rate (BER), and channel capacity for parameters molecular half-life, symbol duration, and communication distance.

\section{Modelling the Molecular Channel}

The complete system considered in this paper consists of a point transmitter, a fully absorbing spherical receiver, a diffusion-based molecular channel, and information-carrying messenger molecules. The molecular channel is a 3-D viscous environment that has infinite volume. The receiver is a spherical surface of radius $\rrn$ with the ability to count the number of absorbed messenger molecules between two instances of time. The transmitter has no volume and is located at distance $r_0$ from the center of the receiver and distance $d$ from the surface of the receiver; hence, $d = r_0 - \rrn$. The messenger molecules are complex compounds that propagate through the molecular channel via Brownian motion. If they degrade, they become useless in terms of the communication (since they do not trigger the receptors). On the other hand, they contribute to the signal when they hit to the receiver boundary/surface prior to degradation.

\subsection{Molecular Signal without Degradation}
\label{subsec:molecularSignalWithoutDegradation}

For the case of no-degradation, we provided the first hitting probability to a spherical absorber in our previous work, \cite{ThreeDimForAbsorbingReceiver}, as
\begin{align}
\label{eqn:pdfNhit}
\pdfNhit =  \displaystyle\frac{\rrn}{r_0}  \frac{r_0 - \rrn}{\sqrt{4 \pi D t^3}}   \exp \left[ - \frac{(r_0 - \rrn)^2}{4Dt} \right].
\end{align}
Moreover, the expected fraction of molecules hitting into the receiver until time $t$ is formulated as
\begin{align}
\label{eqn:cdfNhit}
\begin{split}
\cdfNhit &= \int\limits_{0}^{t} \pdfNhitt{t'}  dt'  \\
         &= \frac{\rrn}{r_0} \text{erfc} \left[\frac{r_0 - \rrn}{\sqrt{4Dt}}\right] ,
\end{split}
\end{align}
which gives the expected number of molecules to be received until time $t$, when multiplied with the number of molecules sent at $t = 0$. From \eqref{eqn:cdfNhit}, we can formulate the expected amount of molecules that are in the channel at each symbol duration. 

\subsection{Incorporating Degradation}
\label{subsec:molecularSignalWithDegradation}

In the biology literature, the degradation of a molecule has been extensively studied as a part of the enzymatic process, which is governed by the chemical reaction
% % % % % % % % % % % % % % % %
\begin{align}
\label{eq:enzym-1}
E + S  \xrightleftharpoons[k_{-1}]{\,k_{1}\,} ES \xrightharpoonup{k_p} E + P ,
\end{align}
% % % % % % % % % % % % % % % %
where $E$, $S$, $ES$, and $P$ denote the enzyme, substrate, enzyme-substrate compound, and the product, respectively. $k_1$, $k_{-1}$, and $k_p$ denote the reaction constants \cite{segel1975enzyme}. The governing differential equations for~\eqref{eq:enzym-1} can be written as
% % % % % % % % % % % 
\begin{align}
\frac{d [S]}{dt} &= - k_1 [E] [S] + k_{-1} [ES] \\ 
\frac{d [E]}{dt} &= - k_1 [E] [S] + k_{-1} [ES] + k_{p}[ES] \label{eq:enzym3} \\ 
\frac{d [ES]}{dt} &= k_1 [E] [S] - k_{-1} [ES] - k_{p}[ES] \label{eq:enzym4} \\ 
\frac{d [P]}{dt} &= k_{p}[ES] \label{eq:enzym5}
\end{align}
% % % % % % % % % % %
where $[\cdot]$ denotes the concentration. In most of the biology literature, the enzymatic reactions are studied under the assumptions where substrate concentration is very high, enzyme concentrations are low enough to halt the first part of the reaction, and the concentration and life-time of enzyme-substrate compound, $ES$, are non-negligible \cite{philippidis1993study, phillips1996stretched}. Under these conditions, the reaction becomes subject to Michaels-Menten dynamics, where the reaction rate is upper limited since the system saturates to a point where no more enzyme is left to react to the remaining substrate. Our communication scenario, however, utilizes a set of assumptions of different nature:
\begin{itemize}
\item The substrate concentration is small since the number of molecules erupted is relatively low (around 1000 molecules per symbol duration).
\item The enzymatic concentration is high since we can control the communication medium as a part of MCvD engineering. 
\item A fast enzyme is utilized, for which the rate constant $k_p$ is high and $k_{-1}$ is low. Following these rate constants, the reaction is one sided, mostly going in the direction of $S \rightharpoonup P$ instead of $S \leftharpoonup P$.
\item The concentration of $ES$ is constant and very low, since $k_p$ is high. Additionally, the small substrate concentration keeps the concentration of $ES$ minimal, limiting the amount of substrate that can form $ES$. As a result, to simplify the equations, hereon we assume $[ES] = \partial_t [ES] = 0$. 
\end{itemize}
Under these conditions, the enzymatic kinetics resemble a perfect catalytic and the governing equations reduce to 
\begin{equation}
\label{eq:enzyme6}
\frac{d [S]}{dt} = - k_1 [E] [S].
\end{equation}
The reasoning is more convincing if one notices $k_{-1}$ is low, which can eliminate the terms $k_{-1}[ES]$ under the listed conditions. Hence, \eqref{eq:enzym3} and \eqref{eq:enzym4} can be reduced to $\partial_t [E] = \partial_t [ES]$, which signifies that the concentration of the enzyme also does not change with time. Here, constants $k_1$ and $[E]$ can be merged into a single coefficient $\lambda$ transforming \eqref{eq:enzyme6} to
\begin{equation}
\label{eq:enzyme7}
\frac{d [S]}{dt} = - \lambda [S],
\end{equation}
which is the first order ordinary differential equation representing exponential decay. 

%In our model, if a messenger molecule degrades before it reaches to the receiver it does not cause a reaction and is not counted as one of the received molecules. Therefore, degradation inevitably reduces the amount of molecules received in a given time frame.

To incorporate molecular degradation into MCvD, we start with the generic exponential decay function
\begin{equation}
\label{eq:exponential_decay}
C(t) = C_0 e^{-t\lambda} ,
\end{equation}
where $C_0$ is the initial concentration, $C(t)$ is the concentration at time $t$, and $\lambda$ is the rate of degradation. While $C_0$ is given as the number of molecules, $\lambda$ is calculated as follows from the corresponding half-life ($\halfLife$) of messenger molecules:
\begin{equation}
\lambda = \frac{\ln(2)}{\halfLife}.
\end{equation}
To find the amount of molecules hitting to the receiver before their decomposition, we employ basic probability of success in events.  If $A$ and $B$ are independent events the probability that event $A$ occurs before event $B$ can be found by
\begin{eqnarray}
\int_0^\infty f_A(t) \cdot P_B(T > t) dt .
\end{eqnarray}
In our case, event $A$ is the arrival of the messenger molecule to the receiver and event $B$ is its degradation. We are interested in the probability that a molecule arrives at the receiver before getting degraded. Therefore, we are dealing with the not-getting-degraded event before the arrival time, and $P_B(T > t)$ corresponds to the event of the degradation time being greater than arrival time $t$. These two events are independent since neither of them affects exponential degradation.\footnote{Notice that, if the events are defined as getting-degraded and reception, events become dependent.} 
The probability of not-getting-degraded for a molecule can be found by the complementary cdf of the exponential distribution as
% % % % % % % % % % % % % % % % % % % %
\begin{align}
P(T \geq t) =  1 - P(T < t) = 1 - (1 - e^{-\lambda t}) = e^{-\lambda t} .
\end{align}
Thus the new channel response function can be easily expressed as
\begin{align}
\label{pdf_decompose}
\pdfDecompose &= \displaystyle\frac{\rrn}{r_0} \frac{r_0 - \rrn}{\sqrt{4 \pi D t^3}}   \exp \left[ - \frac{(r_0 - \rrn)^2}{4Dt} -\lambda t \right] . 
\end{align}
Following \eqref{pdf_decompose}, the combined probability of getting absorbed before getting degraded can be found as follows
\begin{align}
\label{eq:hitBeforeDecompose_early}
\hitBeforeDecompose &= \int_0^\infty \pdfDecompose \, dt \nonumber \\
					&= \int_0^\infty \pdfNhit \, e^{-\lambda t} \, dt .
\end{align}
One can observe \eqref{eq:hitBeforeDecompose_early} is conveniently the Laplace transform of $\pdfNhit$, and yields to
\begin{eqnarray}
\label{hitBeforeDecompose_late}
\hitBeforeDecompose = \frac{\rrn}{r_0} \exp \left[ - \sqrt{\frac{\lambda}{D}} (r_0 - \rrn) \right] .
\end{eqnarray} 
To further detail our problem, we investigate the probability of hitting to the receiver before exponential degradation and an arbitrary time $t$. This is the degradation-enabled counterpart of \eqref{eqn:cdfNhit} and gives the fraction of initial molecules that are absorbed by the receiver till time $t$ for the release at $t = 0$
\begin{align}
\label{hitBeforeDecomposeT}
\begin{split}
\hitBeforeDecomposeT &= \int_0^t \pdfNhit \times e^{-\lambda t} dt  \\
					 &= \hitBeforeDecompose - \frac{\rrn}{2 r_0} \cdot e^{ - \sqrt { \frac{\lambda}{D}} (r_0 - \rrn)}  \\ 
					 &\times \left\lbrace \text{erf} \left(\frac{r_0 \!-\! \rrn}{\sqrt{4Dt}} \!-\! \sqrt{\lambda t} \right) \!+\! e^{2\sqrt{\frac{\lambda}{D}} (r_0 - \rrn)} \right.  \\
					 &\times \left. \left[ \text{erf} \left(\frac{r_0 \!-\! \rrn}{\sqrt{4Dt}} \!+\! \sqrt{\lambda t}\right)- 1 \right] +1 \right\rbrace .
\end{split}
\end{align}

For validation, one can check; first, the $t$ dependent part of \eqref{hitBeforeDecomposeT} expectedly approaches 0 as $t \rightarrow \infty$ and satisfies $\hitBeforeDecomposeAtT{ t \rightarrow \infty} = \hitBeforeDecompose$. Second, it converges to $\cdfNhit$ as $\lambda \rightarrow 0$ or $\halfLife \rightarrow \infty$, thus, meeting the case with no-degradation.

Lastly, to model the expected number of arrivals in a time frame (later a symbol duration), we define the discrete channel response $F_c$ from an arbitrary $t_1$ to $t_2$ as
\begin{equation}
\label{eq_channelResponse}
\channelResponse{t_1}{t_2} \!=\! \hitBeforeDecomposeAtT{t_2} \!-\! \hitBeforeDecomposeAtT{t_1} \,,
\end{equation}
%&=& \frac{\rrn}{2 r_0} \cdot e^{ - \sqrt { \frac{\lambda}{D}} (r_0 - \rrn)} \left\lbrace \text{erf} \left(\frac{r_0 - \rrn}{\sqrt{4Dt_2}} - \sqrt{\lambda t_2} \right) \right. \nonumber \\
%					 &+& \left. e^{2\sqrt{\frac{\lambda}{D}} (r_0 - \rrn)} \left[ \text{erf} \left(\frac{r_0 - \rrn}{\sqrt{4Dt_2}} + \sqrt{\lambda t_2}\right)- 1 \right] +1 \right. \nonumber \\
%					 &-& \left. \left[ \text{erf} \left(\frac{r_0 - \rrn}{\sqrt{4Dt_1}} - \sqrt{\lambda t_1} \right) + e^{2\sqrt{\frac{\lambda}{D}} (r_0 - \rrn)} \left[ \text{erf} \left(\frac{r_0 - \rrn}{\sqrt{4Dt_1}} + \sqrt{\lambda t_1}\right)- 1 \right] +1 \right] \right\rbrace 
which is formally the expected fraction of molecules that arrive in the time interval of $[t_1, t_2]$ for a symbol that was sent at time $t_0$, where $t_0 \leq t_1 < t_2$.
Hence, the expected number of received molecules between $t_1$ and $t_2$, $\nrxts{t_1}{t_2}$, becomes
% % % % % % % % % % % % % % % % % % % % %
\begin{equation}
\label{eq_nrxMean}
\mathbb{E}[\nrxts{t_1}{t_2}] = \ntxs \, \channelResponse{t_1}{t_2} \\
\end{equation}
where $\mathbb{E}[.]$ and $\ntxs$ are the expectation operator and the number of released molecules, respectively. 

% % % % % % % % % % % % % % % % % % % % % % % %
\begin{figure}[t]
\centering
\includegraphics[width=0.55\columnwidth]{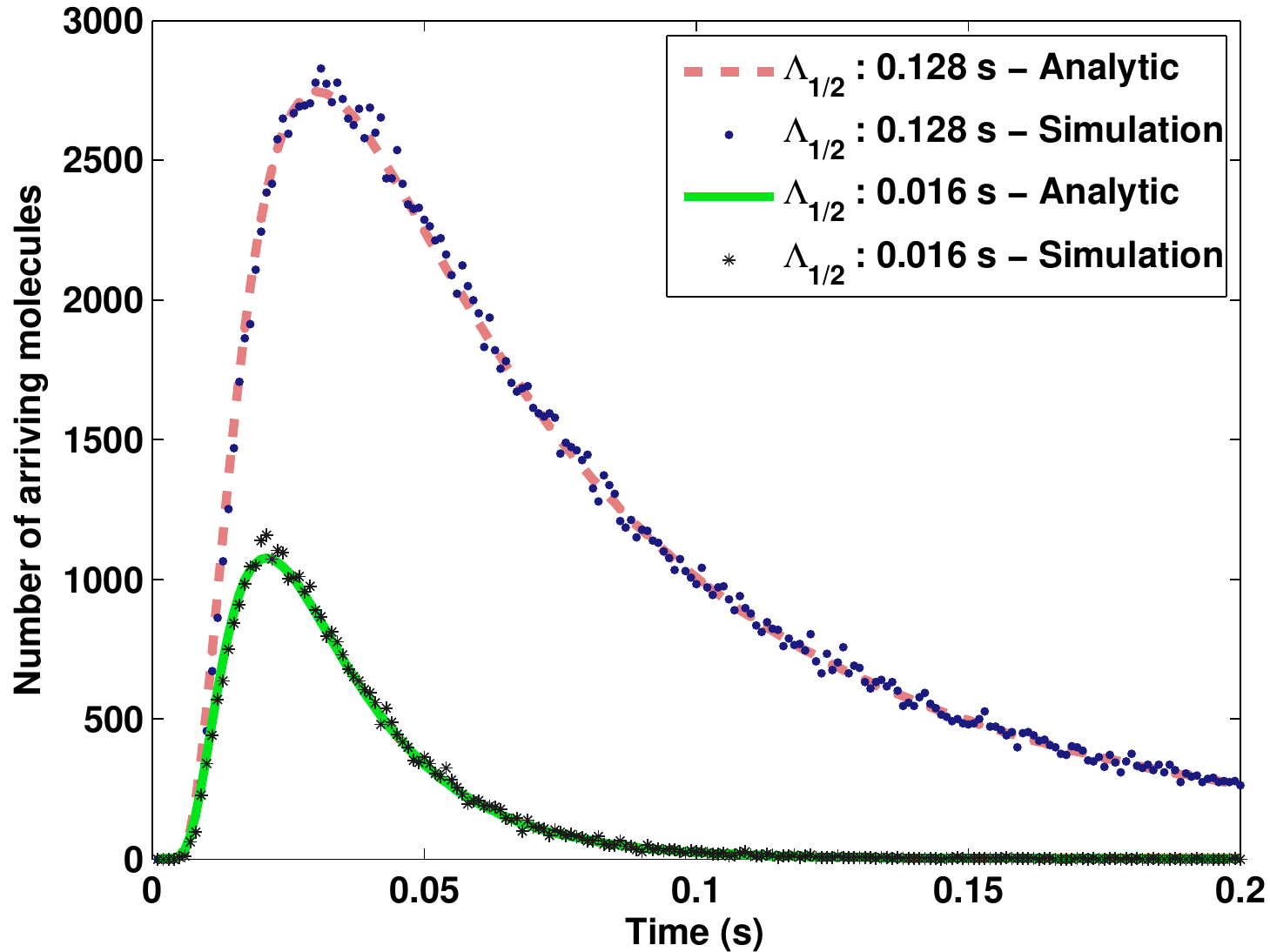}
\caption[Hit time histogram of arrivals]{Hit time histogram of the messenger molecules for an initial release of $\ntxs = 100\,000$ molecules. Analytic solution displays coherence with the simulation results. ($d = 4~\mu m$, $\rrn=10~\mu m$, $D=79.4 ~\mu m^2/s$). }
\label{fig:Analytic_vs_Simulation}
\end{figure}
% % % % % % % % % % % % % % % % %

Figure \ref{fig:Analytic_vs_Simulation} plots the number of arriving molecules with respect to time, and their behavior under two different degradation scenarios with $\halfLifeT{0.016}$ and $\halfLifeT{0.128}$. The figure plots both simulation and analytical results for an initial release of $\ntxs = 100\,000$ molecules. For analytical plots, we evaluate $F_c$ at discrete intervals of $10^{-3}$ seconds (i.e., $\channelResponse{t}{t + 10^{-3}}$ is plotted for every discrete $t$ from 0 to 0.2 seconds with 0.001 second increments). To be consistent and accurate, all the simulations are conducted for the same topology using a step size of $\Delta t = 10^{-6} ~s$. The received molecules are summed and reported for every 1000 step. The simulation results are coherent with the analytical formula with the exception of small deviations.

\subsection{Arrival Modeling}
\label{sec:arrival_modeling}

The formulations derived in Sections~\ref{subsec:molecularSignalWithoutDegradation} and \ref{subsec:molecularSignalWithDegradation} deal with the expected number of molecules arriving to the receiver until time $t$. $\nrxts{0}{t}$ is a random variable and the stochastic nature of the arrival process of diffusion can be modeled via a binomial distribution where the success probability is $\hitBeforeDecomposeT$ or $\cdfNhit$ depending on whether the decomposition of molecules is considered or not. If the decomposition of molecules is not considered, $\nrxts{0}{t}$ is formulated as
% % % % % % % % % % % % % % % % % % % % % %
\begin{align}
\label{eq_nrxRandom}
\nrxts{0}{t} \sim \mathcal{B}(\ntxs , \cdfNhit )
\end{align}
where $\mathcal{B}(n,p)$ is the binomial random variable with $n$ trials and success probability $p$. When we incorporate the decomposition of messenger molecules, the model becomes
% % % % % % % % % % % % % % % % % % % % % %
\begin{align}
\label{eq_nrxRandomWithDecomp}
\nrxts{0}{t} \sim \mathcal{B}(\ntxs , \hitBeforeDecomposeT ).
\end{align}

% % % % % % % % % % % % % % % % %
\begin{figure*}[!t]
        \centering
        \begin{subfigure}[t]{0.5\textwidth}
                        \includegraphics[width=\textwidth]{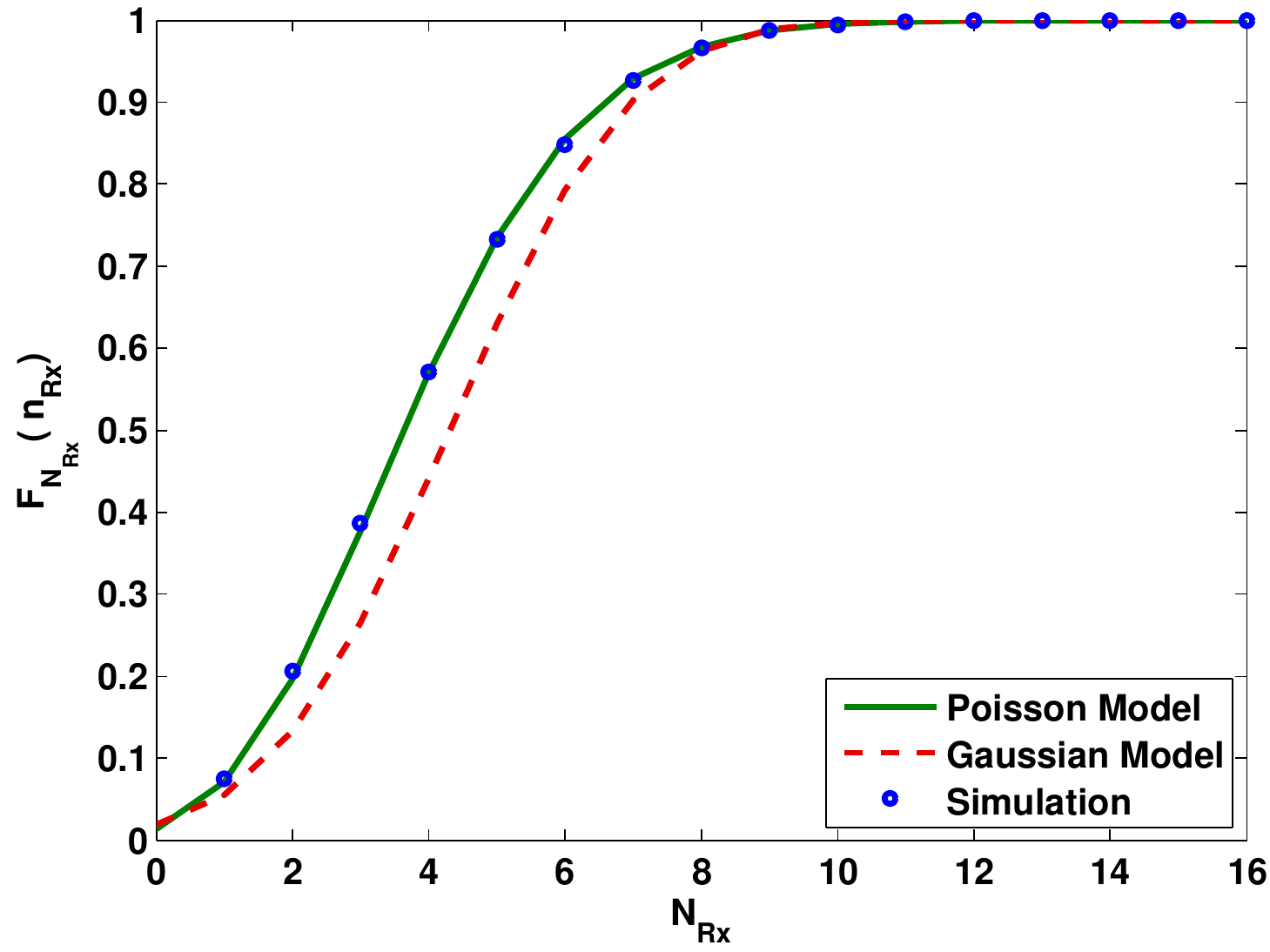}
                        \caption{$t_{\mbox{start}}=4~s$, $t_{\mbox{end}}=4.2~s$}
                        \label{fig:arrivalModelingP}
        \end{subfigure}%
        \begin{subfigure}[t]{0.5\textwidth}
                        \includegraphics[width=\textwidth]{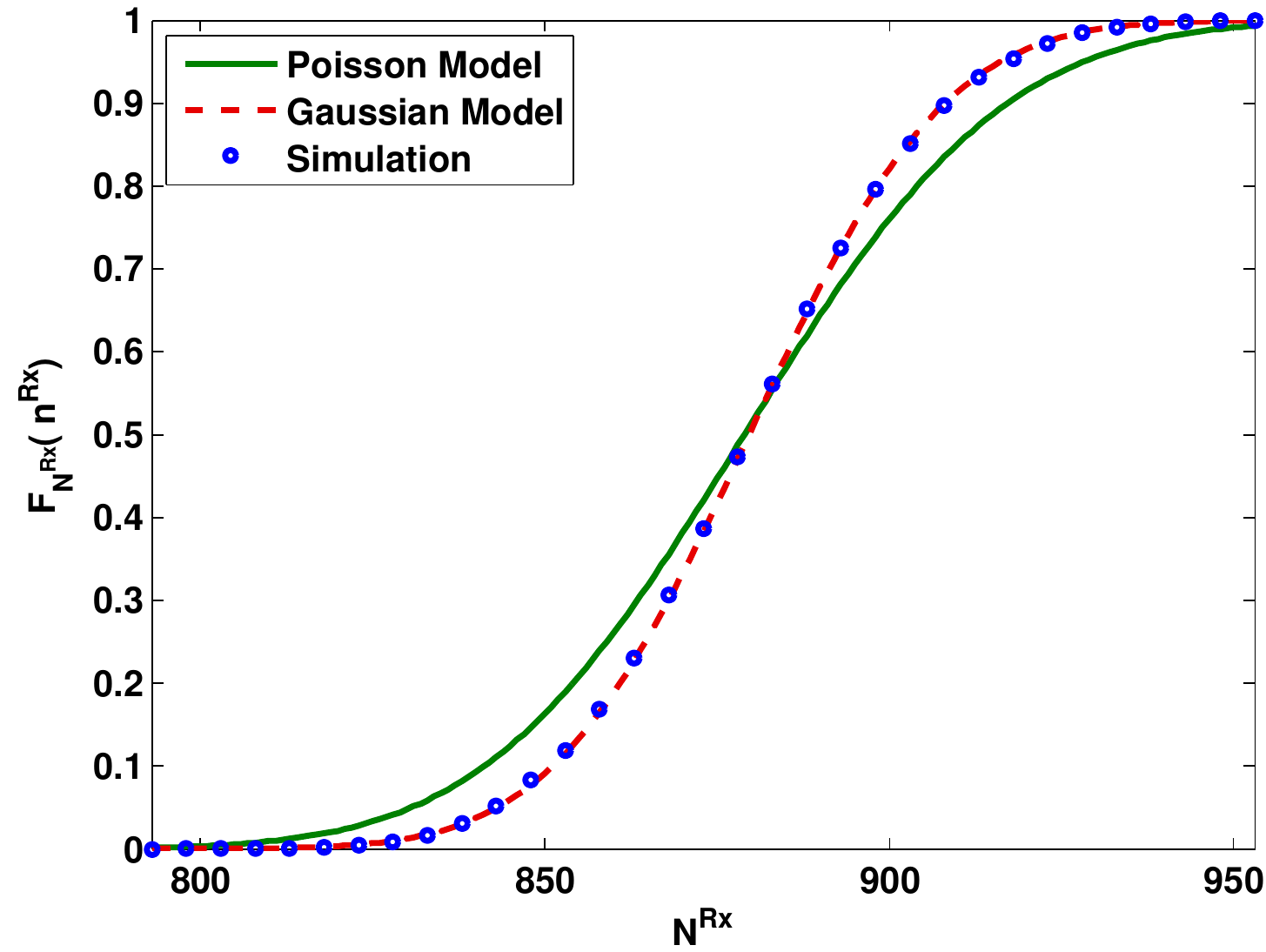}
                        \caption{$t_{\mbox{start}}=0~s$, $t_{\mbox{end}}=0.4~s$}
                        \label{fig:arrivalModelingG}
        \end{subfigure}
        \caption{CDF of $\nrxts{t_{\mbox{start}}}{t_{\mbox{end}}}$. For small number of received molecules Poisson approximation to binomial distribution achieves better compatibility, whereas for large number of received molecules Gaussian approximation gives better results ($N = 2000$, $d = 4~\mu m$, $\rrn=10~\mu m$, $D=79.4 ~\mu m^2/s$, $N_{\mbox{replication}} = 5000$).}
        \label{fig:arrivalModeling}
\end{figure*}
% % % % % % % % % % % % % % % % % % %

With extensive simulations we show that the original process is binomial. It is, however, hard to formulate the error probabilities for consecutive symbols with a binomial model since there is a need to sum the binomial random variables. Therefore, we analyze the approximations for binomial distribution and show that the binomial random variable $\nrxts{0}{t}$ can be approximated by a Poisson or Gaussian random variable, depending on the conditions.

In Figure \ref{fig:arrivalModeling}, the x-axis corresponds to $\nrxts{t_{\mbox{start}}}{t_{\mbox{end}}}$ and the y-axis shows the CDF value. We can easily see which approximation is better under which condition. Poisson approximation is better than Gaussian approximation with rare events~\cite{vonmises1964mathematicalTP}, therefore, a smaller mean number of arriving molecules, creating a rare arrival process, makes the Poisson approximation better. Figure \ref{fig:arrivalModeling} supports the statement since the Poisson random variable approximates the original process better than the Gaussian random variable for smaller values of $\mathbb{E}[\nrxts{t_{\mbox{start}}}{t_{\mbox{end}}}]$. On the contrary, if $\mathbb{E}[\nrxts{t_{\mbox{start}}}{t_{\mbox{end}}}]$ is large enough, then the Gaussian approximation is better than the Poisson approximation. Molecular degradation, inevitably reduces the amount of received molecules and forces the system to operate in smaller detection thresholds for the best performance. Hence, the systems with molecular degradation are better modeled with Poisson approximation, which favors the case where a small number of molecules are received (i.e., rare arrival process). 

\subsection{Modulation \& Demodulation}

In order to understand the effect of molecular degradation, we create a communication scenario, in which time is divided into equal-length time slots at which one bit is sent through the molecular channel. In this scenario, we also assume the transmitter and the receiver are fully synchronized, as explained in~\cite{moore2012synchronization}. 

In this study, we use concentration shift keying (CSK) since, which is the most basic and widely applied technique. In this technique, we modulate the information on the amount of messenger molecules that are emitted in a symbol duration ($\tsym$). This modulation type is analogous to amplitude shift keying (ASK) modulation in electromagnetic (EM) communication and requires a thresholding-based demodulation scheme. 

The most straightforward use of CSK is the binary CSK (BCSK) modulation scheme where two levels of concentration are used to modulate one bit of information. In our scenario we erupt $N_0$ molecules to represent bit-0 and $N_1$ molecules to represent bit-1. To increase the gap between $N_0$ and $N_1$, we choose $N_0 = 0$ and $N_1 = N$, where $N$ is the maximum number of molecules that can safely be released from a cell in a single signal burst.

\begin{figure}[t]
\centering
\includegraphics[width= 0.55\columnwidth]{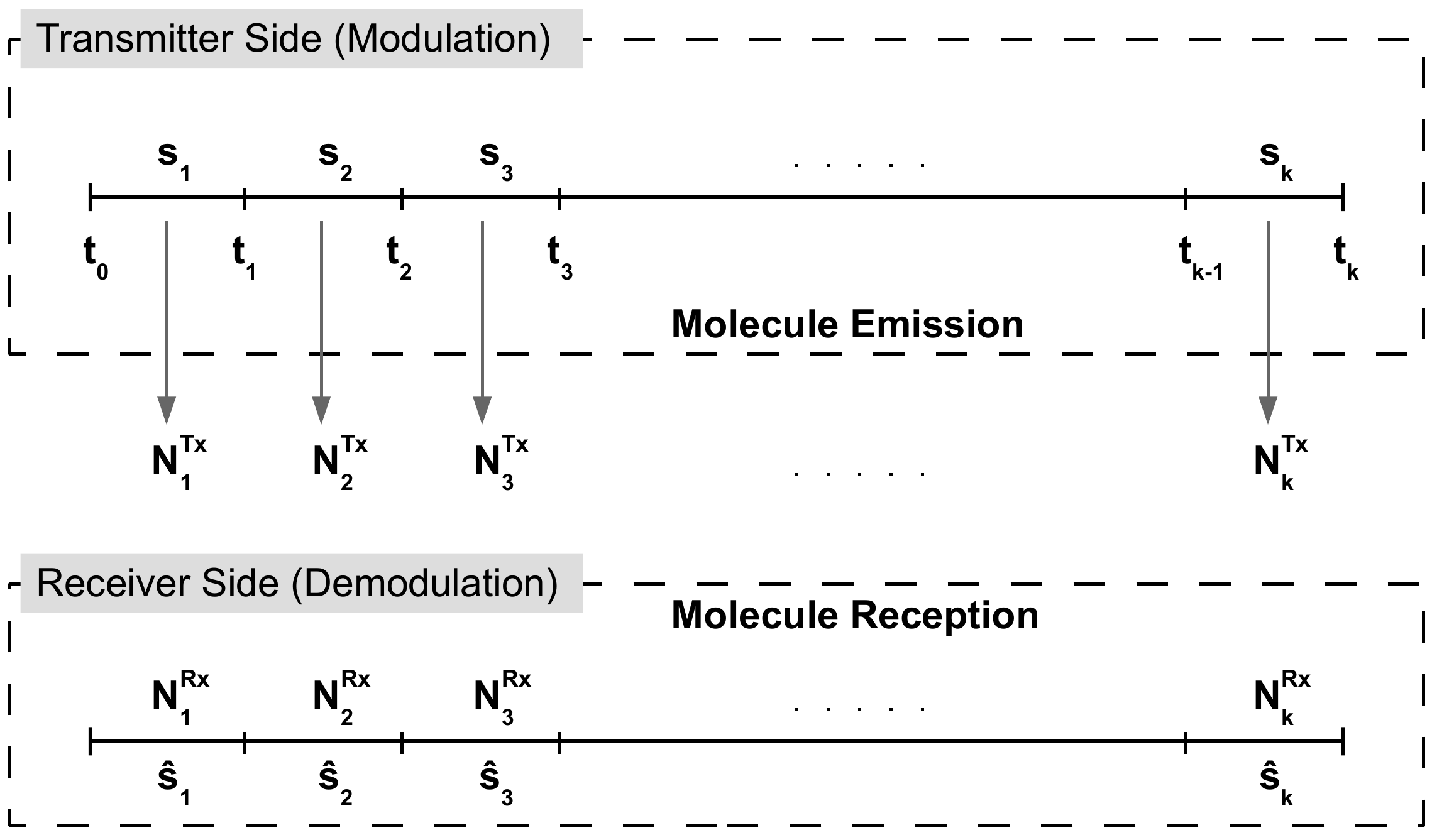}
\caption{A schema of the continuous communication scenario. The transmitter node releases $\ntxsi{i}$ molecules depending on the current bit value, $s_i$. Similarly, the receiver demodulates the amount of absorbed molecules ($\Yts{i}$) in each symbol duration. 
}
\label{fig_modulation_demodulation}
\end{figure}

On the demodulation side, we assume that the receiver counts the number of molecules received for each symbol duration and compares that with a predetermined threshold, $\tau$. As previously described in Section \ref{sec:arrival_modeling}, the number of molecules arriving at the receiver in the $\ith$ symbol duration is a binomially distributed random variable, denoted by $\Yts{i}$. The receiver demodulates $i^{th}$ bit value as ``0'' if $ \Yts{i} \leq \tau $, and as ``1'' otherwise.  
% % % %
\begin{equation}
\label{eq:demodulation_rule}
\decodedsi{i}=\mathcal{D}(\Yts{i}) = \left\{  
						\begin{array}{ll}
						0, & \Yts{i} \leq \tau \\
						1, & \text{otherwise} \,,
						\end{array} 
					\right.
\end{equation}
where $\decodedsi{i}$ and $\mathcal{D}(.)$ represent the demodulated $\ith$ symbol and the demodulation function for the received molecules, respectively. In Figure \ref{fig_modulation_demodulation}, a simple schema of the continuous communication is depicted. In the figure, in each slot, the transmitter releases $\ntxsi{i}$ molecules depending on the bit value, denoted by $s_i$. Analogously the receiver counts the number of received molecules, $\Yts{i}$, in each time slot and decides the intended bit by the rule given by \eqref{eq:demodulation_rule}.
% % % % % % % % % % % % % % % % % % % % % % % % % % % %

\subsection{Symbol Detection Probabilities}

The number of received molecules in one symbol duration is originally binomially distributed as explained in Section~\ref{sec:arrival_modeling}. The parameters of the binomial distribution come from (\ref{hitBeforeDecomposeT}) and from the number of released molecules. We can formulate the correct detection probability of bit-1 when $\symi{i}=1$ as
\begin{equation}
\label{eqn:Pc1_first}
Pc_{|1} (\decodedsi{i}\!=\!1 | \symi{i}\!=\!1, \tau, \lambda) = 1 - Pr(\Yts{i} \leq \tau  | \symi{i}\!=\!1, \tau, \lambda) \,,
\end{equation}
where $\Yts{i}$ is the number of received molecules in the $\ith$ symbol duration. Therefore, if we consider only the molecules released in the $\ith$ symbol duration and discard the effect of previous symbol emissions, the probability of correct detection of bit-1 becomes
\begin{align}
\begin{split}
Pc_{|1} (\decodedsi{i}\!=\!1 | \symi{i}\!=\!1, \tau, \lambda) 
				&= 1- Pr(\Yts{i} \leq \tau | \symi{i}\!=\!1, \tau, \lambda)  \\ 
				&= 1- \sum\limits_{k = 0}^{\tau} \dbinom{N_1}{k} w^k ~(1-w)^{N - k}\, , 
\end{split}
\end{align}
where $N_1$ is the number of molecules sent to represent a bit-1.

In reality, $\Yts{i}$ is affected by the previous emissions, so we have to incorporate all the molecules in the environment that are sent in the previous symbol duration. The new formula for the $\ith$ symbol becomes 
\begin{align}
\begin{split}
Pc_{|1} (\decodedsi{i}\!=\!1 | \symi{i}\!=\!1, \symsq{1}{i-1}, \tau, \lambda) 
		&= 1 - Pr(\Yts{i} \leq \tau | \symi{i}\!=\!1, \symsq{1}{i-1}, \tau, \lambda) \\ 
		&= 1- Pr(\Ytsprev{i}{1}+...+\Ytsprev{i}{i} \leq \tau) \, ,
\end{split}
\end{align}
where $\symsq{1}{k}$ denotes the symbol sequence between 1 and $k$, and $\Ytsprev{i}{m}$ values denote the number of received molecules in the $\ith$ symbol duration due to emission at the $\mth$ symbol duration. $\Ytsprev{i}{m}$'s are also binomially distributed random variables. Note that, we have $N_0=0$ to decrease the energy consumption and maximize the signal separation~\cite{kuran2010energy, kim2013novelMT}. This formula for correct detection of the $\ith$ bit, requires a recursive algorithm to work and is computationally too expensive to solve for large $i$ \cite{butler1993distribution}. For this reason, we employ a well-known  Poisson approximation for the binomial distribution as
% % % % % % % % % % % % % % % % %
\begin{eqnarray}
\mathcal{B}(n, p) \sim \mathcal{P}(np) .
\end{eqnarray}
% % % % % % % % % % % % % % % % % 
This approximation is known to be successful in the case of large $n$ and small $p$, which are in our problem the number of molecules sent in each symbol duration and the fraction of arrivals. With this approximation, we can derive a closed form solution for the correct detection probability of bit-1 in the $\ith$ symbol duration as
% % % % % % % % % % % % % % % % %
\begin{equation}
Pc_{|1} (\decodedsi{i}\!=\!1, |\symi{i}=1, \symsq{1}{i-1}, \tau, \lambda, N_1) = 1 - \left( e^{-N_1\, \channelResponseSumFromTo{1}{i}} ~\sum\limits_{k = 0}^{\tau} \frac{[N_1  \channelResponseSumFromTo{1}{i}]^k}{k!} \right) ,
\end{equation}
% % % % % % % % % % % % % % % % %
where $N_1$ is the number of molecules sent to represent bit-1 throughout the communication, and $ \channelResponseSumFromTo{1}{i}$ is the sum of all channel responses that represent bit-1 until the $\ith$ symbol and the current symbol itself. Since we modulate bit-0 with the emission of zero molecules, we just sum the emissions due to bit-1. $\channelResponseSumFromTo{1}{i}$ can be calculated as
\begin{eqnarray}
\channelResponseSumFromTo{1}{i} = \sum\limits_{k = 1}^{i} \symi{k}\,\channelResponse{t_{k-1}}{t_{k}}  \, ,
\end{eqnarray}
where $t_{m}$ is the point in time where $\mth$ symbol duration ends, i.e., $t_i = i \, \tsym$ where $\tsym$ is the symbol duration, and $\symi{m}$ is the bit value of the $\mth$ bit.

Given this calculation for $Pc_{|1}$, one can easily derive $Pc_{|0}$ for the $\ith$ symbol as 
\begin{equation}
Pc_{|0} (\decodedsi{i}\!=\!0, |\symi{i}=0, \symsq{1}{i-1}, \tau, \lambda, N_1) = e^{-N_1 \channelResponseSumFromTo{1}{i}} ~\sum\limits_{k = 0}^{\tau} \frac{[N_1 \, \channelResponseSumFromTo{1}{i}]^k}{k!} .
\end{equation}
Here, to derive the overall error probabilities for the channel, we calculate the average of all errors presented thus far and conclude
\begin{equation} \label{eq:pc0}
Pc_{|0} (\decodedSRV \!=\!0| S\!=\!0, \tau, \lambda, N_1 ) = \lim\limits_{i \rightarrow \infty} \sum\limits_{z = 1}^{i} \frac{Pc_{|0} (\decodedsi{z}\!=\!0 | \symi{z}=0, \symsq{1}{z-1}, \tau, \lambda, N_1)}{i} ,
\end{equation}
and
\begin{equation} \label{eq:pc1}
Pc_{|1} (\decodedSRV \!=\!1| S\!=\!1, \tau, \lambda, N_1 ) = \lim\limits_{i \rightarrow \infty} \sum\limits_{z = 1}^{i} \frac{Pc_{|1} (\decodedsi{z}\!=\!1 | \symi{z}=1, \symsq{1}{z-1}, \tau, \lambda, N_1)}{i},
\end{equation}
% % % % % % %
where $\decodedSRV$ and $S$ denote the random variable of decoded and intended symbol. Finally from \eqref{eq:pc0} and \eqref{eq:pc1}, we calculate the error probabilities as
\begin{equation} \label{eq:pe0}
Pe_{|0} (\decodedSRV\!=\!1| S\!=\!0, \tau, \lambda, N_1 ) = 1 - Pc_{|0} (\decodedSRV \!=\!0| S\!=\!0, \tau, \lambda, N_1 ) ,
\end{equation}
\begin{equation} \label{eq:pe1}
Pe_{|1} (\decodedSRV\!=\!0| S\!=\!1, \tau, \lambda, N_1 ) = 1 - Pc_{|1} (\decodedSRV \!=\!1| S\!=\!1, \tau, \lambda, N_1 ) ,
\end{equation}
and the overall probability of error as
\begin{align}
\begin{split}
Pe &= Pe_{|0} (\decodedSRV\!=\!1| S\!=\!0, \tau, \lambda, N_1 ) \,\pi_0\\
   &+ Pe_{|1} (\decodedSRV\!=\!0| S\!=\!1, \tau, \lambda, N_1 ) \,\pi_1,
\end{split}
\end{align}
where $\pi_0$ and $\pi_1$ denote probability of sending bit-0 and bit-1. Note that in the further sections of the paper, we simplify the notation to improve readability. For this purpose, respective errors for bit-0 and bit-1 are denoted with $\pe[0]$ and $\pe[1]$.

% % % % % % % % % % % % % %
\begin{figure}[th]
\centering
\includegraphics[width=0.55\columnwidth]{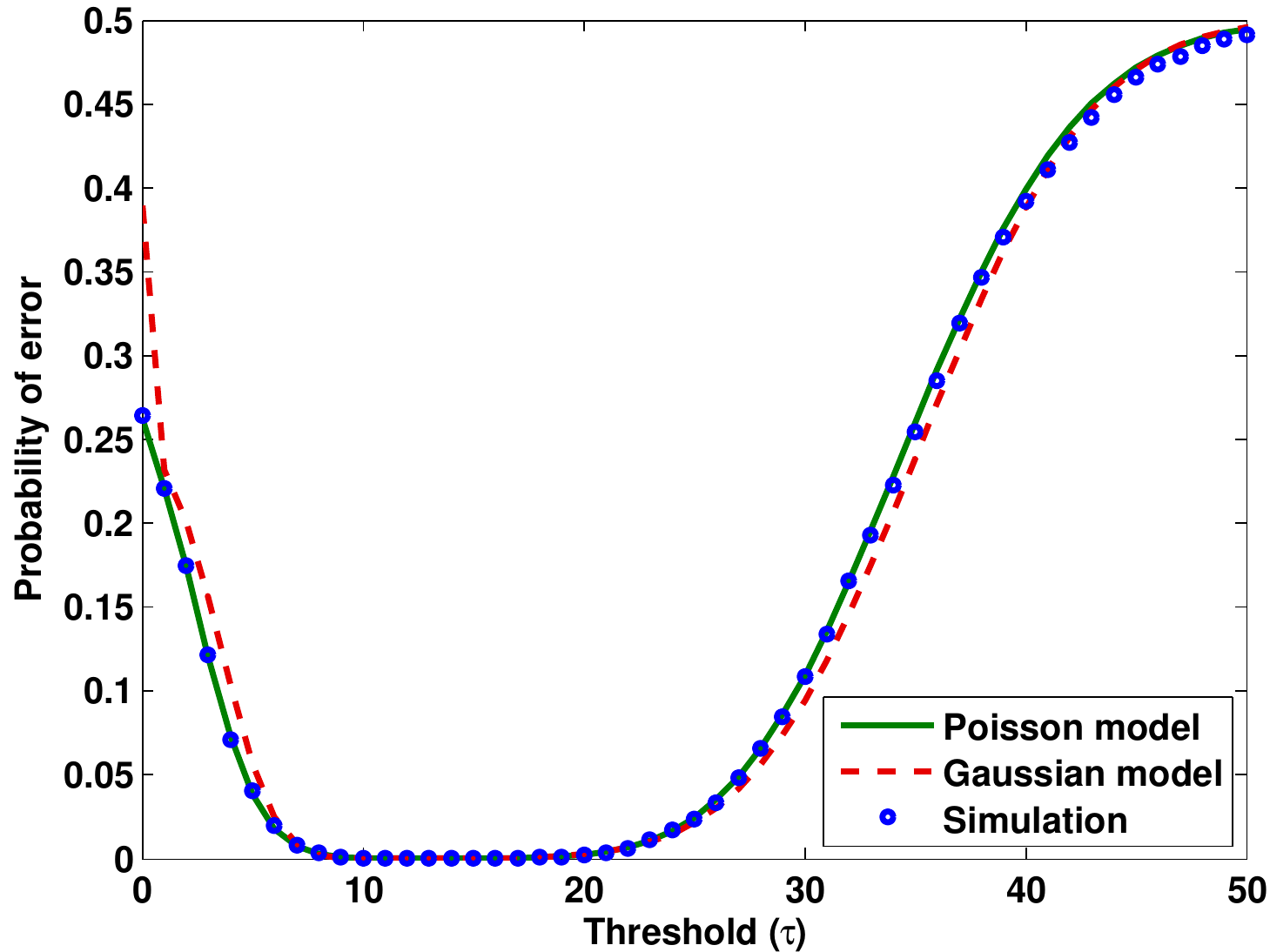}
\caption{Probability of error ($\perr$) for various thresholds ($\tau$). We observe a clear similarity between the simulation results and Poisson model presented through \eqref{eqn:Pc1_first} to \eqref{eq:pc1}, ($\tsym = 0.06 ~s$, $d = 4 ~\mu m$, $N_1 = 1000$, $D = 79.4 ~\mu m^2/s$, $N_0 = 0$, $\halfLifeT{0.016}$, and $\pi_0=\pi_1=0.5$).
}
\label{fig:Pc1Pc0_poiss_binom}
\end{figure}
% % % % % % % % % % % % % % % % %

The diffusion channel is significantly affected by the long tail of the arrival distribution of the molecules. The number of arriving molecules in each symbol duration is determined by the collective sum of arrivals from all previously sent symbols in the channel and their order of appearance. Consequently, the detection of an arbitrary symbol in a communication scenario is heavily influenced by the preceding symbols. This is why \eqref{eq:pc0} and \eqref{eq:pc1} have no closed form and can only be numerically evaluated. Fortunately, in such a channel, \eqref{eq:pc0} and \eqref{eq:pc1} experience a fast convergence to an accurate value. 

In Figure \ref{fig:Pc1Pc0_poiss_binom}, we present the probability of error for various threshold values. The figure indicates clear agreement between the simulation results and the Poisson model, whilst the Gaussian model, abundantly found in literature \cite{kilinc2013receiver, mahfuz2013strength}, displays a higher margin of error. The reason behind is that the degradation lowers the expected number of arriving molecules and the Gaussian model is error prone in small numbers. Concerning the molecular channel, we observe that the optimal threshold values are between $[10, 20]$ where $\perr$ nearly drops down to 0. As we go left, we observe an incline as a result of increased $\pe[0]$ since smaller threshold values tend to favor bit-1. Similarly, the right tail of the curve is influenced by $\pe[1]$ where larger thresholds result in an  increased number of missed-detections.

\section{Characteristics of the Molecular Channel with Molecular Degradation}

In diffusion-based communication, due to the probabilistic nature of the Brownian motion of messenger molecules, signal shaping is hard to achieve. With the introduction of molecular degradation, however, we shape the MCvD signal into a more desirable form with fewer stray molecules and earlier peak time. In this section, we first elaborate on the effects of degradation onto the shape of the signal by investigating the changes on pulse peak time and pulse peak amplitude. Second, we define a new metric, interference-to-total-received-molecule ratio (ITR), to quantify ISI and show how radical the effect of degradation on ISI is under various scenarios. 

\subsection{Pulse Peak Time}
As shown in Figure~\ref{fig:Analytic_vs_Simulation}, an MCvD signal has one peak. Hence we can find the mean pulse peak time, $\pulsept$, by finding the vanishing point for the derivative of $\pdfDecompose$ with respect to time
% % % % % % % % % % % % %
\begin{equation}
\frac{\partial\pdfDecompose}{\partial t}  = \partial_t \left( \frac{\rrn}{r_0} \frac{d}{\dortPiDe} \expoPiDe e^{-\lambda t} \right) = 0.
\label{eqn:pulsePeakTime}
\end{equation}
% % % % % % % % % % % % %
While calculating $\partial_t \pdfDecompose$, we arrive at one crucial intermediate step
% % %
\begin{equation}
\frac{\rrn}{r_0} \frac{d}{\sqrt{4 \pi D}} \left\lbrace t^{-3/2} \, e^{ - \frac{d^2}{4 D t} - \lambda t} \left( - \frac{3}{2t} + \frac{d^2}{4D t^2} - \lambda \right) \right\rbrace = 0.
\end{equation}
Solving for $t$ requires
\begin{equation}
\label{eq:pulsePeakBeforeRevision}
- \frac{3}{2t} + \frac{d^2}{4D t^2} - \lambda = 0 .
\end{equation}
Revising \eqref{eq:pulsePeakBeforeRevision} gives us the quadratic equation
\begin{equation}
\label{eq:pulsePeakAfterRevision}
4 D \lambda t^2 + 6 D t - d^2 = 0 .
\end{equation}
At this point, one can evaluate \eqref{eq:pulsePeakAfterRevision} in two ways. First, if we consider the regular diffusion process where $\lambda = 0$, the coefficient of $t^2$ vanishes, leading to 
\begin{equation}
\mathbb{E}[\pulsept] = \frac{d^2}{6D}.
\label{eqn:pulsePeakTimeNoDegradation}
\end{equation}
Second, if there is molecular degradation and $\lambda \neq 0$, the solution for $\pulsept$ becomes
\begin{equation}
\label{eq:t_peak_degradation}
\mathbb{E}[\pulsept | \lambda] = \frac{\sqrt{36D^2 + 16 D d^2 \lambda} - 6D}{8D \lambda}.
\end{equation}
%%%%%%%%%%%%%%%%%%%%%%%%%%%%%%%%%%%%%

\begin{figure}[t]
\centering
\includegraphics[width= 0.55\columnwidth]{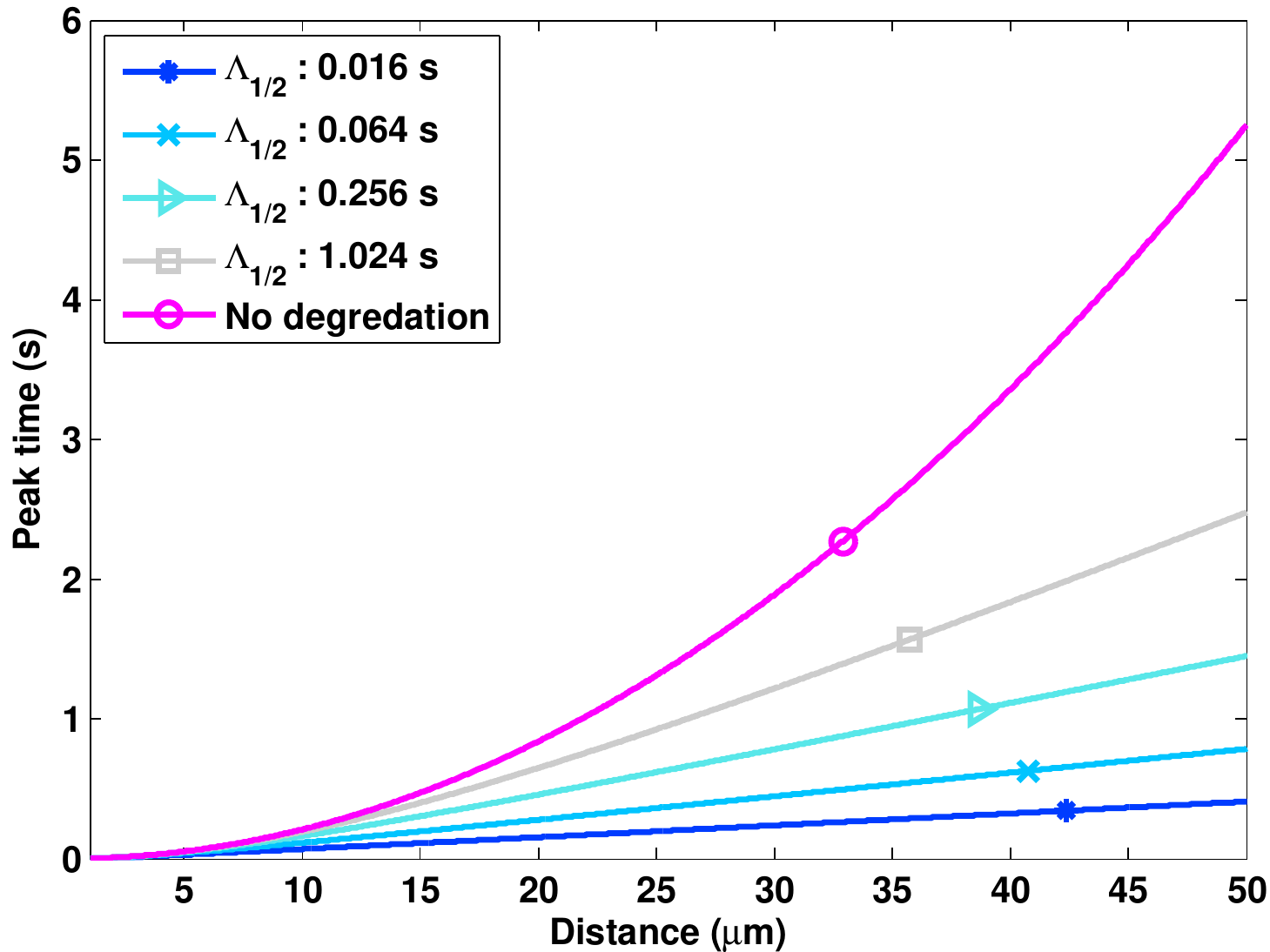}
\caption[Distance versus peak time of the signal]{Distance versus peak time of the signal. With degradation, peak time and distance correlation is linear instead of quadratic ($\rrn=10~\mu m$, $D=79.4 ~\mu m^2/s$). }
\label{fig:t_peak}
\end{figure}

%%%%%%%%%%%%%%%%%%%%%%%%%%%%%%%%%%%%%
In EM communications, $\pulsept$ is proportional to the propagation time, which is the distance divided by the wave propagation speed, hence it is proportional to $d$. In the molecular communication system without degradation, due to the diffusion dynamics, $\pulsept$ is proportional to $d^2$. With the introduction of the degradation (even if the degradation process is very slow), however, this proportion again reduces to $d$. Therefore, with the introduction of degradation, we can passively shape the molecular signal and reduce the restricting effect of diffusion dynamics on the signal travel time. 

In Figure \ref{fig:t_peak}, the distance versus $\mathbb{E}[\pulsept]$ is depicted for different degradation scenarios. In the figure, we can observe that, without degradation, the change in the peak time with respect to distance is quadratic, whereas it is linear when messenger molecules undergo degradation. 

\subsection{Pulse Amplitude}

The pulse amplitude is the maximum amount of arriving molecules in a fixed-size frame of time. $\pulsepn$ can be considered as the amount of molecules left after the channel attenuation, which is highly dependent on the communication distance. 

Since the molecule arrivals to the absorbing receiver is a counting process,\footnote{The diffusion process should be evaluated differently for the non-absorbing (concentration-based) receiver. In the concentration-based receiver, we can conclusively decide the amount of molecules that are in the receiver zone at any instance of time, which nullifies the requirement for a small time window. Nonetheless, although it may seem easier to model, the non-absorbing case is not a satisfactory model since it diverges receiver dynamics from reality. For example, $\pulsepn$ would not depend on the diffusion coefficient, $D$, contrary to the absorbing receiver case. The diffusion coefficient, however, has a significant impact that can be determined experimentally \cite{redner2001guide, berg1993random}. In addition, most of the MCvD receivers found in the nature are absorbing or behave as one.} we cannot deal with the number of arriving molecules at a single point in time. Instead, we must define an interval for which the number of arrivals is highest among all other intervals of the same length. To do this, we start with $\pulsept$ where the highest amplitude is expected. Then, we take the channel response between a time frame of $-\xi/2$ and $+\xi/2$ from $\pulsept$. When we formulate $\pulsepn$, we find
% % % % % % % % % % % % %
\begin{align}
\mathbb{E}[\pulsepn] &= \ntxs \channelResponse{\pulsept - \xi/2}{\pulsept + \xi/2},
\end{align}
and to easily manipulate it we use the definition of $F$ given in \eqref{eqn:cdfNhit} and consider it as a Riemann sum to find
\begin{align}
\begin{split}
\mathbb{E}[\pulsepn] &= \ntxs \int\limits_{\pulsept - \xi/2}^{\pulsept + \xi/2}\,  \pdfDecomposeAtT{\pulsept} dt \\
					&\approx \, \ntxs \, \xi \,\pdfDecomposeAtT{\pulsept}
\end{split}
\label{eq:pulsePeakValue}
\end{align}
% % % % % % % % % % % % %
where $\ntxs$ is the number of molecules that are initially released. 
The value of $\pulsepn$ in this case depends on the receiver node's radius $\rrn$, distance $d$, degradation constant $\lambda$ and diffusion coefficient $D$. Note that as $\xi \rightarrow 0^+$, the approximate value of $\mathbb{E}[\pulsepn]$ defined in  \eqref{eq:pulsePeakValue} becomes more accurate.

At this point we again need to separate cases with and without degradation. For the latter case, the calculation of $\mathbb{E}[\pulsepn]$ is neater, the end result yielding to
\begin{equation}
\mathbb{E}[\pulsepn] \approx \ntxs \,\xi \, \frac{\rrn}{d+\rrn} \frac{D}{d^2} \frac{e^{-3/2}}{\sqrt{\pi/54}} .
\end{equation}

For a fixed $\rrn$, we have $\pulsepn \sim 1/d^3$, and this behavior reveals the difference compared to EM communications. If we ignore fading, the amplitude of EM pulse propagating in free space decreases proportional to the square of the transmission distance. The amplitude of a pulse in the MCvD channel, however,  decreases proportional to the cube of the distance. 

Secondly, we calculate the peak amplitude of the degradation scenario using \eqref{eq:t_peak_degradation}, which yields to the rather complicated equation
\begin{align}
\begin{split}
\mathbb{E}[\pulsepn | \lambda] &\approx \ntxs \,\xi \,  \displaystyle\frac{\rrn}{\rrn + d} \frac{d}{\sqrt{4 \pi D \, (\tpeakdegradation)^3}}  \\
&\times \exp \left[ - \frac{2\lambda d^2}{\sqrt{36D^2 + 16 D d^2 \lambda} - 6D} \right. - \left. \frac{\sqrt{36D^2 + 16 D d^2 \lambda} - 6D}{8D} \right], 
\end{split}
\end{align}
where, this time, for a given $\rrn$ we have $\pulsepn \sim \frac{1}{d^{3/2} \, e^d}$ while $\lambda$, being the coefficient of $d$, determines the steepness of the path-loss curve.

%%%%%%%%%%%%%%%%%%%%%%%%%%%%%%%%%%%%%
\begin{figure}[tbp]
\centering
\includegraphics[width= 0.55\columnwidth]{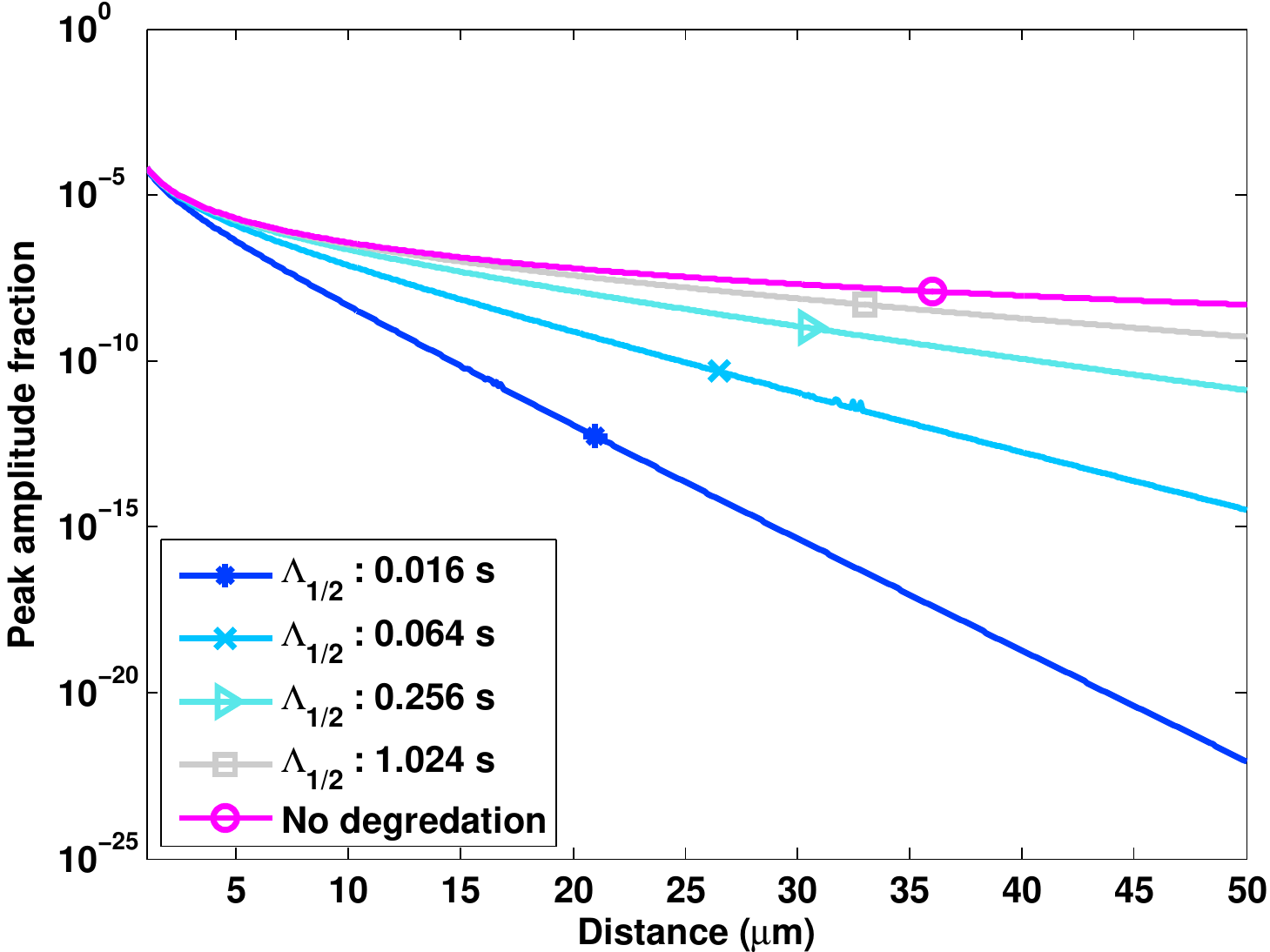}
\caption{Distance versus peak amplitude of the signal for $ \xi = 10^{-6} s$. With degradation, the path loss for peak amplitude becomes exponential instead of polynomial. ($\ntxs = 1$ to represent a fractional value, $\rrn=10~\mu m$, $D=79.4 ~\mu m^2/s$). }
\label{fig:n_peak}
\end{figure}

%%%%%%%%%%%%%%%%%%%%%%%%%%%%%%%%%%%%%

%%%%%%%%%%%%%%%%%%%%%%%%%%%%%%%%%%%%%%%%%%%%%%%%%%%%%%%%%%%%%%%%%%%%%%%%%%%%%%%%
\begin{table*}[th!]
\caption{Diffusion channel characteristics comparison matrix.}
\label{tab:comparison}
\renewcommand{\arraystretch}{1.3}
\centering
\begin{tabular}{ccccc}
\hline  Metric   & Physical Relation & Electromagnetic   & MCvD \cite{ThreeDimForAbsorbingReceiver}& MCvD with degradation (Proposed) \\
\hline  
        $\pulsept$ & Propagation Time      & $d$     & $d^2$	 & $d$ \\
        $\pulsepn$ & Path Loss      & $1/d^2$   & $1/d^3$ & $\frac{1}{d^{3/2} \, e^d}$\\
\hline
\end{tabular}
\end{table*}
%%%%%%%%%%%%%%%%%%%%%%%%%%%%%%%%%%%%%%%%%%%%%%%%%%%%%%%%%%%%%%%%%%%%%%%%%%%%%%%%

Table~\ref{tab:comparison} summarizes the comparison for propagation time and path loss. In terms of channel capacity, MCvD without degradation has a clear disadvantage as, first, the propagation time of the information is proportional to the square of the distance and second, the path loss is one order of magnitude larger. Introducing degradation to the MCvD system seems to mitigate the issue with the propagation time, at the cost of an exponential decrease in peak amplitude with respect to distance. This indicates long distance communications would not favor high degradation rates, while short distance ones benefit from it.  

Figure \ref{fig:n_peak} depicts the distance versus peak pulse amplitude fraction of $\ntxs$ for various degradation scenarios. We observe the rate of decrease in fast degradation cases is much higher than those with slow or no-degradation. At this point we might argue the choice of degradation rate. Clearly, the $\halfLifeT{0.016}$ case would not work well for a distance of 50 micrometers, since detection at the receiver site would not be easy. However, if we choose a degradation level of $\halfLifeT{1.024}$, we would gain twice in propagation speed while losing 10 times as many molecules compared to the no-degradation case. Although this seems like a non-favorable scenario, the energy consumption of producing 10 times more molecules would be trivial for an animal cell, since the energy cost of messenger molecule creation is significantly low \cite{kuran2010energy}. 

To conclude, introducing messenger molecule degradation to an MCvD system allows the designer of a nano-network to shape the molecular signal in the propagation environment. This signal shaping especially benefits the channel in mitigating ISI and providing a more clear signal for the receiver. We will elaborate on this in more detail in Section \ref{sec:performance_evaluation}.

\subsection{Interference to Total Received Molecule Ratio (ITR)}

\begin{figure}[t]
\centering
\includegraphics[width= 0.55\columnwidth]{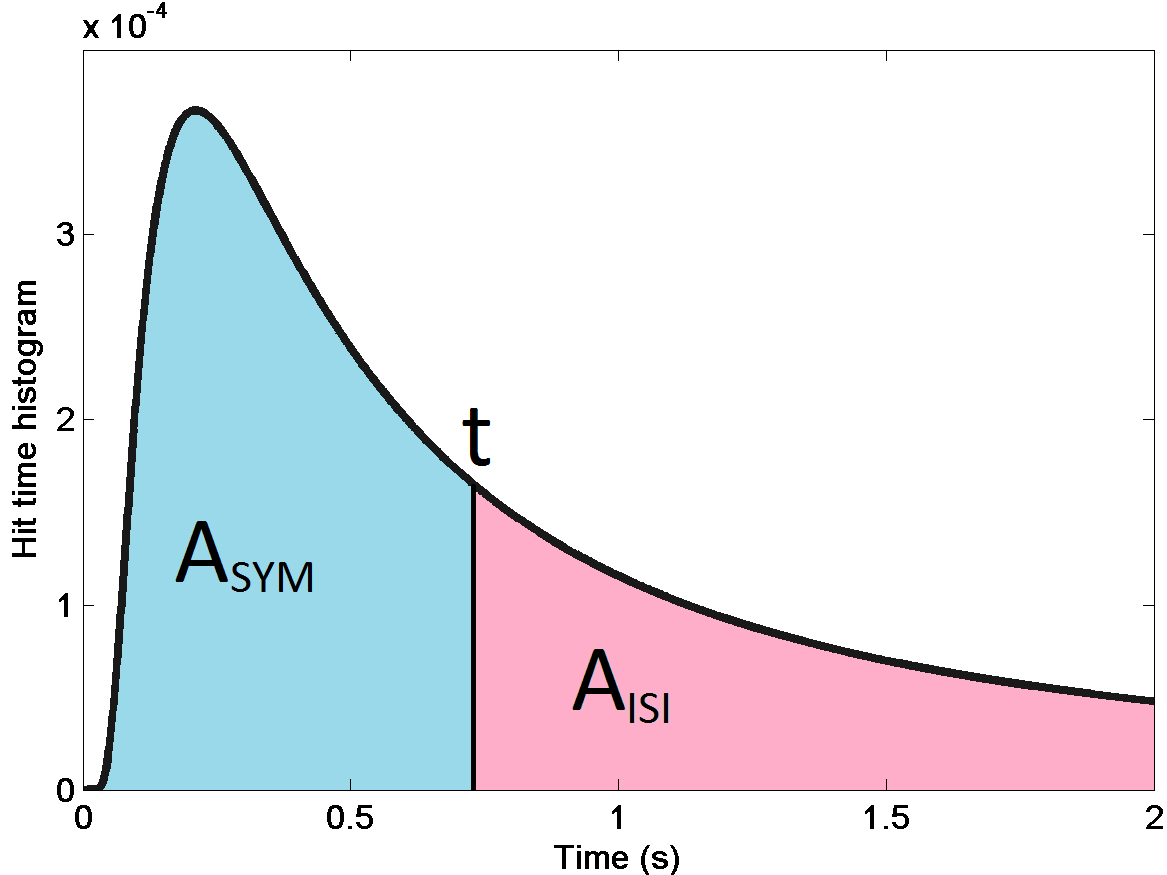}
\caption{Representation of signal power versus interference power depending on time $t$. Molecules that are inside the interference area ($\text{A}_{\text{ISI}}$) result in erroneous detection for upcoming symbol durations.}
\label{fig_interference_to_total_energy_ratio}
\end{figure}

ISI is one of the biggest causes of communication impairment in MCvD systems. Stray molecules from previous symbol durations pile up and impair the correct reception ability of the receiver. Utilizing degradation helps removing these molecules, resulting in better system performance. In order to formulate the amount of molecules causing ISI in the system we create a new metric, $\rateOfISI{t}$, that denotes the fraction of molecules remain unreceived for one burst of molecules. We can formulate it with $\text{A}_{\text{ISI}}/(\text{A}_{\text{ISI}}+\text{A}_{\text{SYM}})$ using the areas depicted in Figure \ref{fig_interference_to_total_energy_ratio}. $\rateOfISI{t}$ is formulated as follows in terms of the channel response functions
% % % % % % % % % % % % % % % % % % % % %
\begin{eqnarray}
\label{N_arr_early}
\rateOfISI{t} = 1 - \frac{\hitBeforeDecomposeT}{\hitBeforeDecomposeAtT{ t \rightarrow \infty}} . 
\end{eqnarray}
Using \eqref{hitBeforeDecompose_late}, \eqref{hitBeforeDecomposeT} and \eqref{N_arr_early} we can come up with the expression
% % % % % % % % % % % % % % % % % % % % % % % % % % % % %
\begin{align}
\label{eqn:rateOfISI}
\begin{split}
\rateOfISI{t} &= \frac{1}{2} \left\lbrace \text{erf} \left(\frac{r_0 - \rrn}{\sqrt{4Dt}} - \sqrt{\lambda t} \right) + e^{2\sqrt{\frac{\lambda}{D}} (r_0 - \rrn)}  \right.  \\
&\times \left[ \text{erf} \left(\frac{r_0 - \rrn}{\sqrt{4Dt}} + \sqrt{\lambda t}\right)- 1 \right] + 1 \bigg\} 
\end{split}
\end{align}
for the fraction of molecules causing ISI. Notice that \eqref{eqn:rateOfISI} is independent of $r_0$ and $r_r$ individually, but dependent on $d = r_0 - r_r$, the distance between the transmitter and the receiver. Therefore, we claim that observations derived from $\rateOfISI{t}$ metric are valid regardless of the receiver size as long as the distance is fixed. 

% % % % % % % % % % % % % % % % % % %
\begin{figure}[t]
\centering
\includegraphics[width= 0.55\columnwidth]{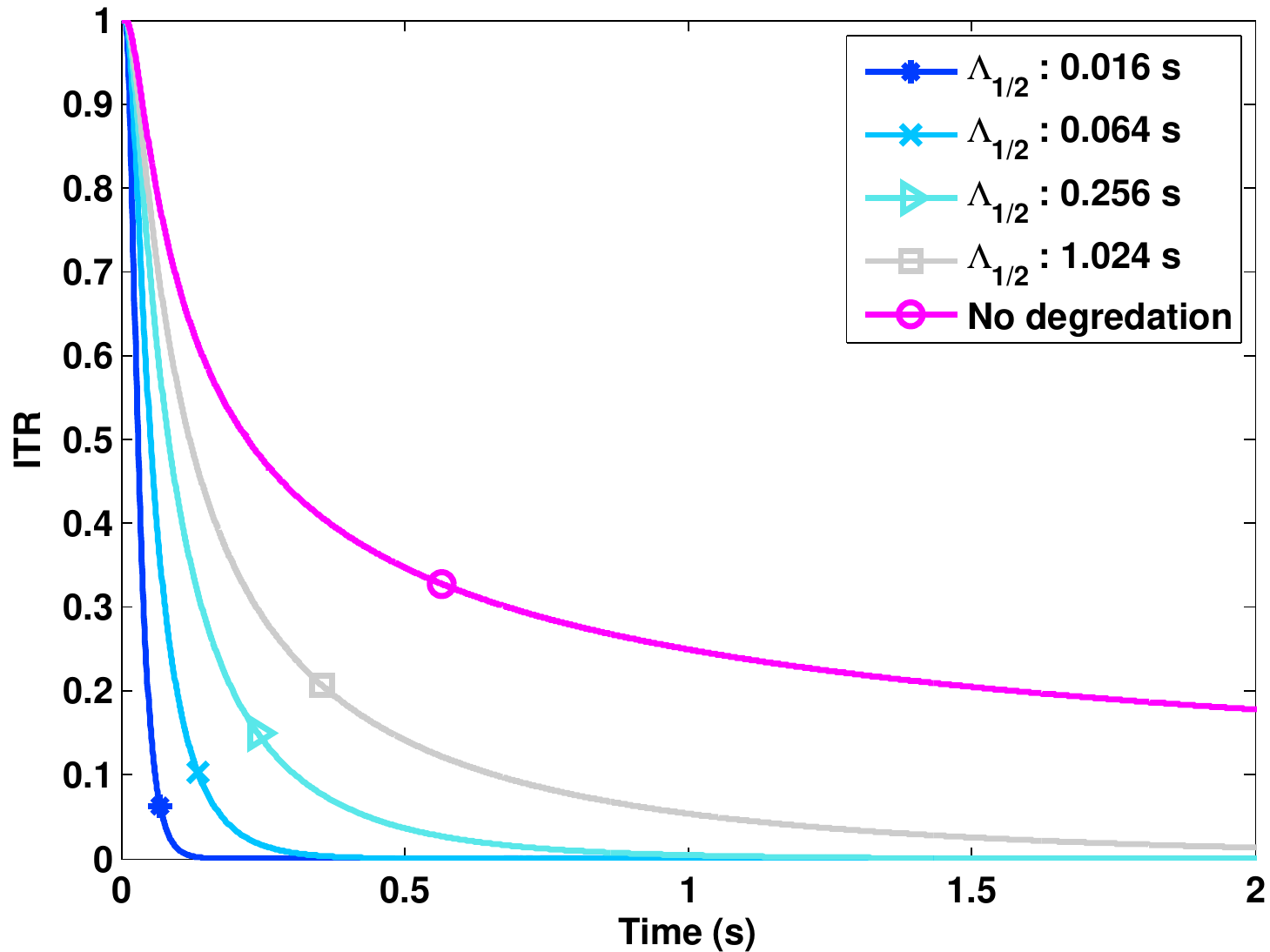}
\caption{
Fraction of molecules causing ISI in the system for various $\halfLife$ values. Faster degradation significantly shortens the tail of the unreceived molecule distribution, decreasing the amount of stray molecules and, consequently, ISI ($d = 4~\mu m$, $\rrn=10~\mu m$, $D=79.4 ~\mu m^2/s$).
%Fraction of molecules causing ISI. Faster degradation clears stray molecules and reduces the amount of ISI in the system.
}
\label{fig:Rate_of_arrivals}
\end{figure}

Figure \ref{fig:Rate_of_arrivals} depicts $ \rateOfISI{t} $ for various half-life values, where faster degradation clearly reduces the amount of stray molecules for a given molecular communication scenario. As observed from the figure, utilization of molecular degradation of half-life of 16 milliseconds clears almost all the stray molecules in the system at $t = 200$ milliseconds. In the no-degradation case, however, more than half of the initially released molecules are still in the system, resulting in significant ISI. 

A second observation from Figure \ref{fig:Rate_of_arrivals} is that, the rate of decline in $\rateOfISI{t}$ reduces as larger half-life (i.e., smaller $\lambda$) values are selected. This reduction is the result of the terms inside error functions in \eqref{eqn:rateOfISI}, where $\lambda$ is the coefficient of $t$.

\section{Performance Evaluation}
\label{sec:performance_evaluation}

\subsection{System Parameters}

The experiments in this paper have a wide range of parameters affecting system behavior. For performance evaluations, we fix some of these parameters to observe the effects of others. To give a clearer vision to the reader, in Table \ref{tb:system_parameters}, we present all the possible parameters that can be investigated for their respective impacts. The table gives the ranges in which the parameters are considered in at least one of the experiments presented through the paper. 
%\told{Notice that some parameters like diffusion coefficient are kept constant even though they affect the system behavior as observed from the fundamental equations. However, in order to keep the dimension of results small, we chose to not further investigate the effects of these parameters and focus on more crucial aspects in communication design.} \tbirkan{@@Birkan(yumusatalim)} 

%\begin{table}[h]
%\begin{center}
%\caption{Range of parameters used in the experiments}
%\renewcommand{\arraystretch}{1.0}
%\label{tb:system_parameters}
%\begin{tabular}{p{5cm} l}
%\hline
%\bfseries{Parameter} 							& \bfseries{Value} \\ 
%\hline 
% Number of messenger molecules ($\ntxs$) 				& $\{1000,\, %100\,000\}$ \\ 
% Liquid viscosity 					& $ 0.001 ~ kg / (s \, m )$ \\ 
% Messenger molecule radius 							& $ 2.56 \cdot 10^{-9} ~m$ \\ 
% Temperature 						& 310 $^{\circ}$K \\ 
% Diffusion coefficient (D) 		& $79.4 ~ (\mu m)^2 / s $ \\ 
% Half-life for messenger molecules ($\halfLife$) 			& $\{0.0005 - 1.024\}~s$ \\ 
% Receiver radius ($\rrn$)               & $10 ~\mu m$                \\
%Transmitter distance to the center of the receiver ($r_0$) & $\rrn + \{1 - 50\} ~\mu m$\\ 
% Detection threshold ($\tau$) 		& $\{1 - \ntxs\}$ \\
% Simulation step size ($\Delta t$)	&  $10^{-6} ~s$  \\
% Symbol duration ($\tsym$)			& $\{0.001 - 1\}~s$\\
% \hline
%\end{tabular} 
%\end{center}
%\end{table}

\begin{table}[!t]
\begin{center}
\caption{Range of parameters used in the experiments}
\renewcommand{\arraystretch}{1.0}
\label{tb:system_parameters}
\begin{tabular}{p{7.2cm} l}
\hline
\bfseries{Parameter} 							& \bfseries{Value} \\ 
\hline 
 Number of messenger molecules ($\ntxs$) 				& $\{1000,\, 100\,000\}$ \\ 
 Liquid viscosity 					& $ 0.001 ~ kg / (s \, m )$ \\ 
 Messenger molecule radius 							& $ 2.56 \cdot 10^{-9} ~m$ \\ 
 Temperature 						& 310 $^{\circ}$K \\ 
 Diffusion coefficient (D) 		& $79.4 ~ (\mu m)^2 / s $ \\ 
 Half-life for messenger molecules ($\halfLife$) 			& $\{0.0005 - 1.024\}~s$ \\ 
 Receiver radius ($\rrn$)               & $10 ~\mu m$                \\
Transmitter distance to the center of the receiver ($r_0$) & $\rrn + \{1 - 50\} ~\mu m$\\ 
 Detection threshold ($\tau$) 		& $\{0 - \ntxs\}$ \\
 Simulation step size ($\Delta t$)	&  $10^{-6} ~s$  \\
 Symbol duration ($\tsym$)			& $\{0.001 - 1\}~s$\\
 \hline
\end{tabular} 
\end{center}
\end{table}

\subsection{Effect of Degradation on Detection Performance}

\begin{figure*}[t]
        \centering
        \begin{subfigure}[t]{0.5\textwidth}
                \includegraphics[width=\textwidth]{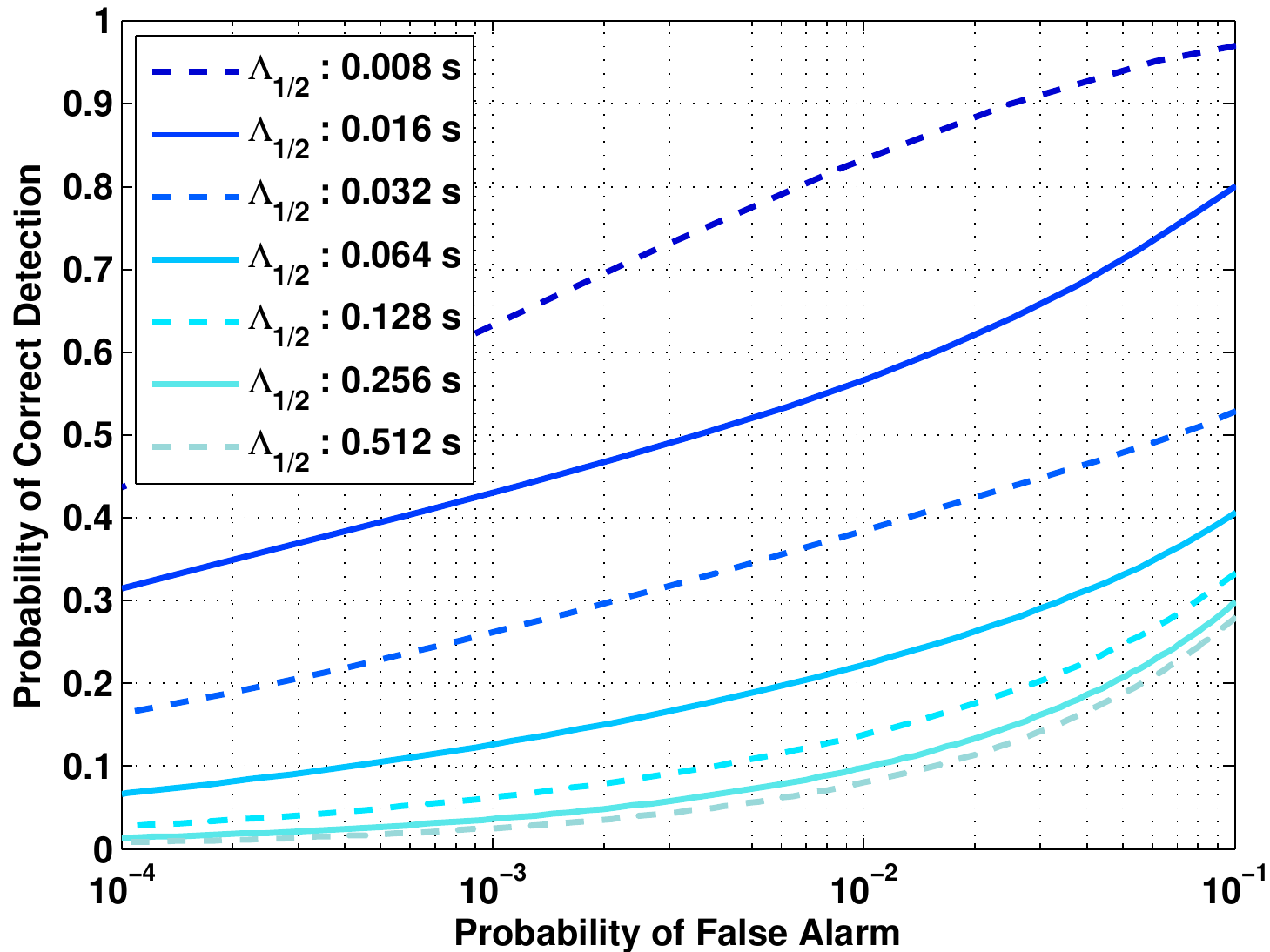}
                \caption{$\tsym = 30~msec$}
                \label{fig:ROC4}
        \end{subfigure}%
        ~ %add desired spacing between images, e. g. ~, \quad, \qquad etc.
        \begin{subfigure}[t]{0.5\textwidth}
                \includegraphics[width=\textwidth]{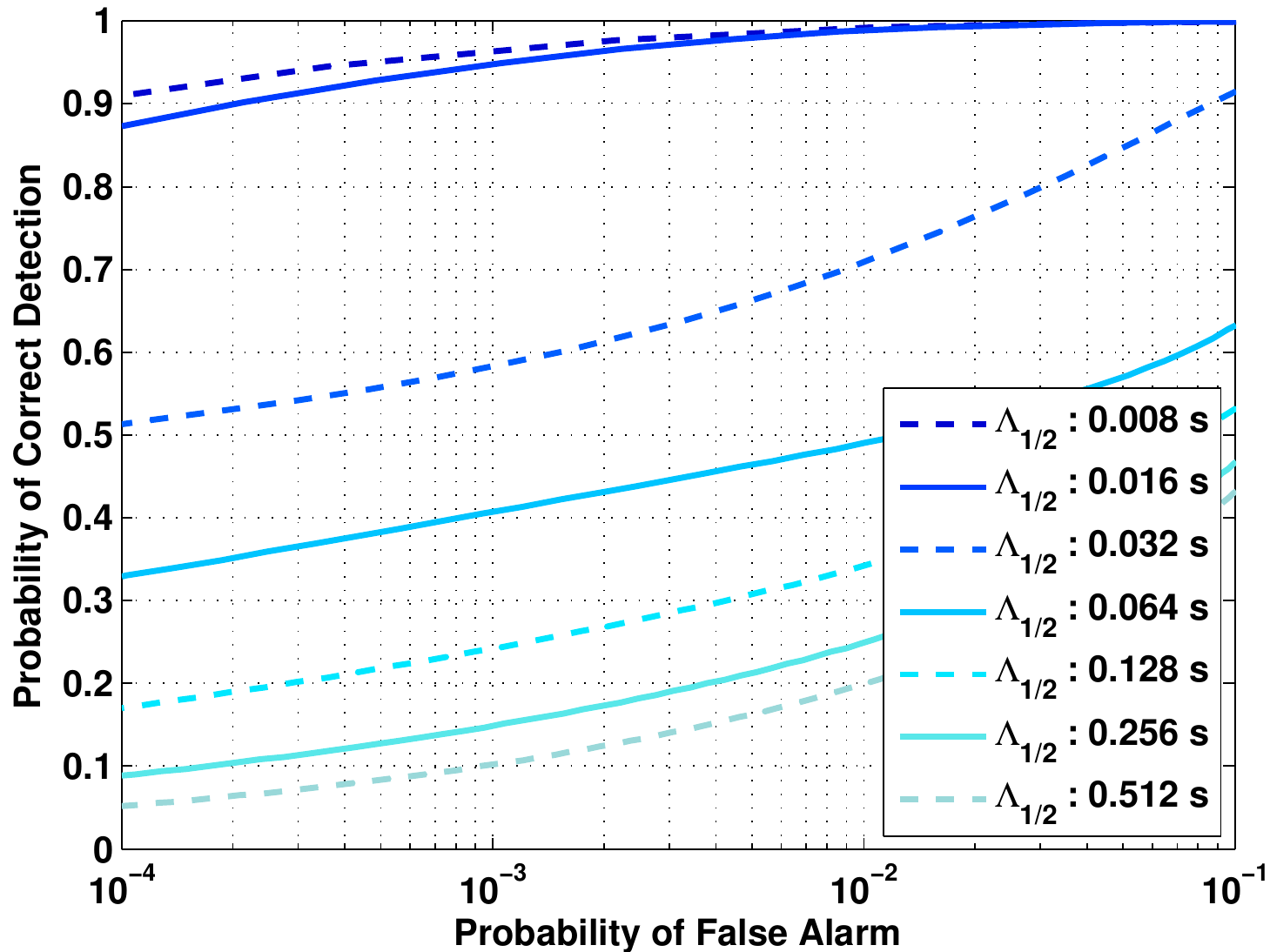}
                \caption{$\tsym = 40~msec$}
                \label{fig:ROC6}
        \end{subfigure}
        %~ %add desired spacing between images, e. g. ~, \quad, \qquad etc.
        %\begin{subfigure}[t]{0.3\textwidth}
        %        \includegraphics[width=\textwidth]{./Graphs/ROC_8}
        %        \caption{A mouse}
        %        \label{fig:mouse}
        %\end{subfigure}
        \caption{ROC curves for various degradation speeds. Systems with higher rates of degradation perform better over systems with lower rates of degradation. An increase in the symbol duration also has a positive effect on the performance of the channel, for every value of half-life we observe an elevated curve in (\ref{fig:ROC6}) ($d = 4~\mu m$, $\ntxs = 1000$, and $\pi_0=\pi_1=0.5$). }
        \label{fig:ROCs}
\end{figure*}

For a reliable communication system, operating with high probabilities of correct detection at the receiver side is preferred. Following the utilization of BCSK, we calculate the probability of false alarm ($P_f$) from \eqref{eq:pe0} where $P_f = 1 - \pc[0] = \pe[0]$, and the probability of correct detection ($P_d$) from \eqref{eq:pc1} where $P_d = \pc[1]$. In Figure \ref{fig:ROCs}, we present two ROC curves with symbol durations equal to 0.03 and 0.04 seconds. 
We observe a compromise between $P_f$ and $P_d$ as usual. This is straightforward if one thinks about varying threshold used for signal detection. If the detection threshold is lowered, it is always possible to achieve better detection performance of bit-1. Reducing $\tau$, however, may result in an incorrect demodulation of bit-0, since stray molecules from previous symbol durations continue to arrive at the receiver, which in turn increases $P_f$. 
% % % % % % % % % % % % % % % % % % % % % % % % % % %

From both of Figures \ref{fig:ROC4} and \ref{fig:ROC6}, we observe a negative correlation between molecular half-life and correct detection performance. Both of the cases show clear improvements on $P_d$ with respect to $P_f$ as the molecular half-lives decrease. This indicates that the degradation of messenger molecules reduces the amount of stray molecules causing ISI in the receiver and increases detection performance. 

Comparing Figure \ref{fig:ROC4} and \ref{fig:ROC6}, we deduce that increased symbol duration also has a positive effect on the ROC curves. When the symbol duration is longer, all cases with different half-lives show better detection performance, where the improvement on $\halfLifeT{0.016}$ case is larger than $\halfLifeT{0.008}$ case. This is also consistently observed in Figure \ref{fig:berPlot}, where $\halfLifeT{0.016}$ makes a larger improvement between symbol durations [0.03, 0.04] and consequently has a larger slope. The reason behind this difference lies within the difference between the two channels' potential for improvement for the selected range of symbol durations. Both systems show very little improvement on $\pe[0]$ values since most of the stray molecules are cleared with each case of degradation speed. This leaves improvement in $\pe[1]$ as the reason for the difference in the potential, where $\halfLifeT{0.008}$ shows a very small amount of improvement, whilst $\halfLifeT{0.016}$ displays a larger one. This potential difference comes from the amount of molecules staying in the system as stray molecules. One can observe, when symbol duration is larger, that these molecules are no longer stray molecules but sources of information transfer to the receiver and result in a reduction in $\pe[1]$. To conclude, while destroying molecules improves ISI, it also reduces the amount of information transferred to the receiver side, limiting the potential for improvement in the system. This limitation causes a significant increase in $\pe[1]$ when we select very small half-lives in the system.

\subsection{Effect of Degradation on Bit Error Rate}
\label{sec:Effect of Degradation on Bit Error Rate}

To observe the effect of the molecular degradation on the error rate in the system, we look at the bit error rate (BER) metric, defined as the minimum total probability of error given a fixed symbol duration, number of molecules and half-life, 
\begin{equation}
\text{BER} = \perr^* = \min\limits_{\tau} \left[ \perr (N, \tau, \lambda, \tsym) \right] .
\end{equation}
% % % % % % % % % % % % % % % % % % %
\begin{figure}[t]
\centering
\includegraphics[width=0.55\columnwidth]{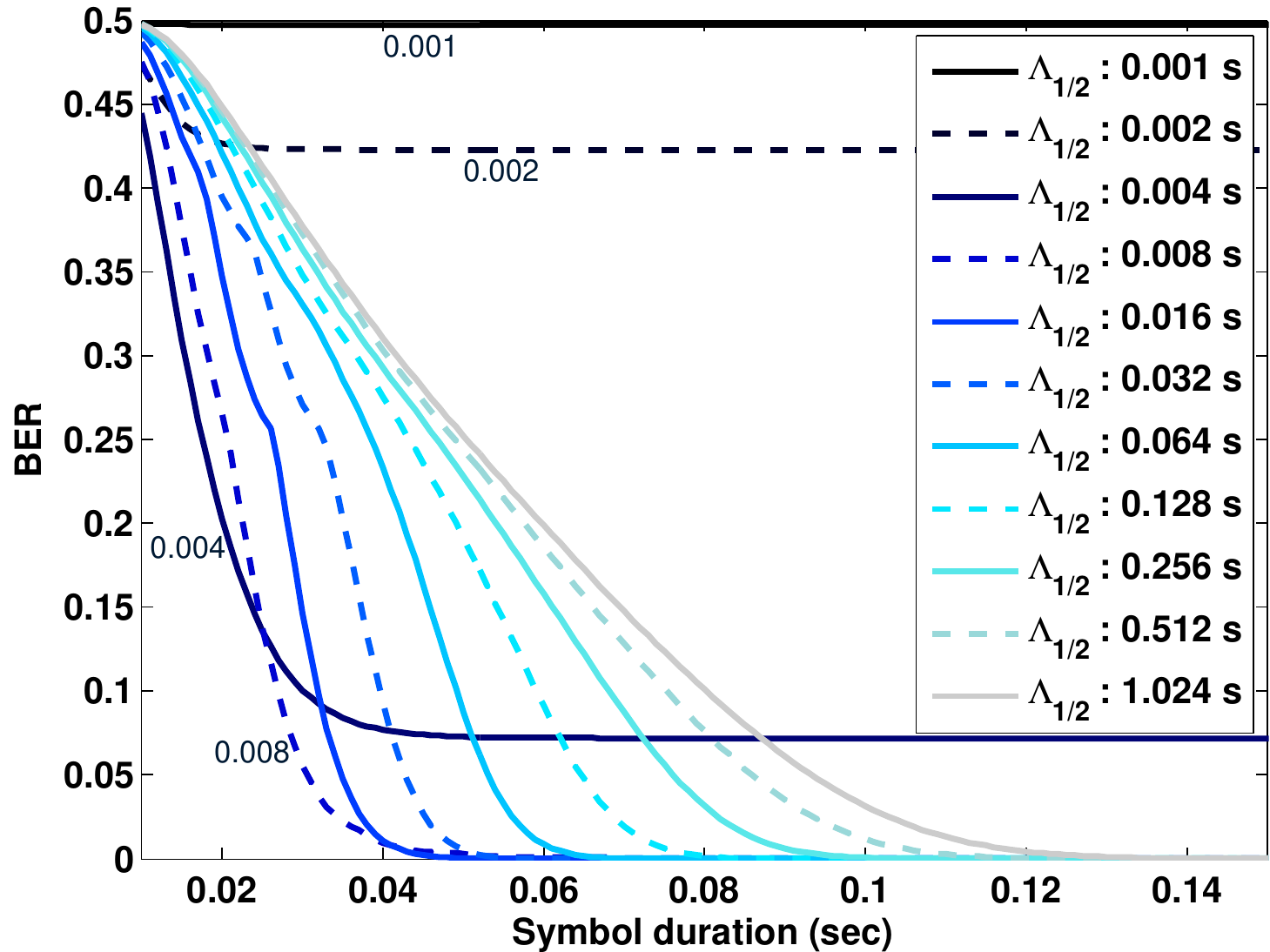}
\caption{BER for various $\halfLife$ values. The rate of decrease in BER increases as shorter half-lives are utilized. Channels with too short half-lives, however, suffer from increased $\pe[1]$, which creates a BER floor (${d = 4~\mu m}$, ${\ntxs = 1000}$).}
\label{fig:berPlot}
\end{figure}
% % % % % % % % % % % % % % % % % % %
In Figure \ref{fig:berPlot}, we show the change in BER for various half-life values as the selected symbol duration increases. One can observe from the figure that down to $\halfLifeT{0.008}$ increases in degradation speed increases the rate of decrease of BER as larger symbol durations are selected. This allows selecting smaller symbol durations without any performance impairment. On the other hand, selecting smaller half-lives than $0.008 s$ results in a BER that has a lower bound caused by the contribution of $\pe[1]$. At larger degradation rates, $\pe[0]$ is the bigger contributor to the overall error since molecular communication systems suffer from ISI significantly. At smaller rates, however, molecules get destroyed so fast that they cannot deliver the encoded information to the receiver. In Figure \ref{fig:berPlot}, we observe $\halfLifeT{0.004}$ has a better BER than $\halfLifeT{0.008}$ for shorter symbol durations. As the symbol durations get longer, however, BER of $\halfLifeT{0.004}$ gets lower bounded and is outperformed by $\halfLifeT{0.008}$. The same condition is also observed for $\halfLifeT{0.002}$ and $\halfLifeT{0.001}$ where they are outperformed much more quickly as they impair correct detection performance more significantly.

\subsection{Effect of Degradation on Channel Capacity }

Capacity is a chief indicator of the performance of a communication system. In this section, we elaborate on the effects of degradation on channel capacity, calculated as
\begin{align}
C^* = \sup\limits_{\tau, t_s, p_S(s)} I(S;\decodedSRV) = \sup\limits_{\tau, t_s, p_S(s)} \,\,\sum\limits_{\decodedsi{} \in \{0, 1\}} \sum\limits_{\symi{} \in \{0,1\}} p(\symi{},\decodedsi{}) \log_2 \left( \frac{p(\symi{},\decodedsi{})}{p_S(\symi{}) p_{\decodedSRV{}}(\decodedsi{})} \right)
\label{eqn_channel_capacity}
\end{align}
where $I(.\,;.)$ stands for mutual information between random variables of intended and decoded symbols. Bayesian rule with $\pc[1]$ and $\pc[0]$ are utilized for evaluating marginal probabilities. For a given system parameters such as distance, diffusion coefficient, and radius of the receiver, we define $t_s$ dependent channel capacity as
\begin{align}
C(t_s) = \sup\limits_{\tau, p_S(s)} I(S;\decodedSRV)
\label{eqn_channel_capacity_ts}
\end{align}
by removing the maximization over $t_s$. Note that, instead of finding channel capacity per channel use we prefer to use channel capacity in bits per second ($bps$), which is calculated as
\begin{align}
C [bps] = C/t_s \,\,.
\label{eqn_capacity_bps}
\end{align}

The outcomes of selecting different degradation rates subject to different symbol durations and communication distances will be detailed in the following subsections via evaluating the channel capacity.

\subsubsection{Symbol Duration Analysis}
\begin{figure}[t]
\centering
\includegraphics[width=0.55\columnwidth]{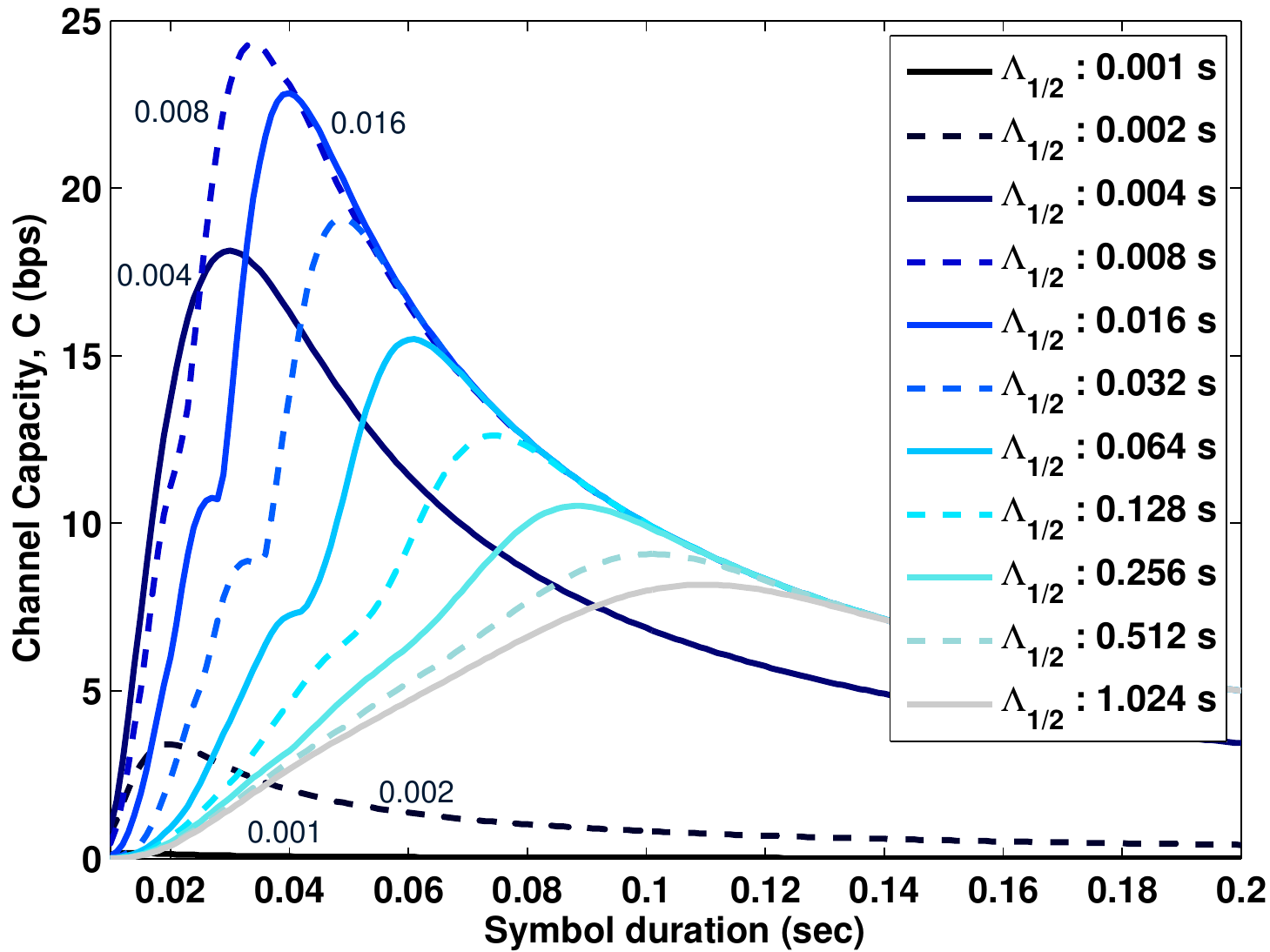}
\caption{
Capacity versus symbol duration for various half-lives. Reducing degradation half-life until $\halfLifeT{0.008}$ increases capacity due to the reduced ISI, however, shorter degradation half-lives lower the capacity in general. For the given parameters, there exist an optimum degradation rate and symbol duration. If the degradation rate is not a degree of freedom, then the optimum symbol duration can be calculated for a given molecule ($d = 4~\mu m$).
}
\label{fig:optimizeTs}
\end{figure}

In Section \ref{sec:Effect of Degradation on Bit Error Rate}, we observe that we can select shorter symbol durations without causing significant increase in $\perr$ when we utilize molecular degradation in a communication via diffusion system. In this section, we elaborate on the selection of the symbol duration length to achieve the maximum capacity.

In Figure \ref{fig:berPlot}, we observe that faster degradation enables the increase of capacity by providing a wider range of symbol durations in which $\perr$ is low. Therefore, in Figure \ref{fig:optimizeTs}, the peak capacity values are at shorter symbol durations for faster degrading molecules, excluding $\halfLifeT{0.001,~0.002,~0.004}$. Consistent with Figure \ref{fig:berPlot}, $\halfLifeT{0.001}$ and $\halfLifeT{0.002}$ systems remain at lower capacity levels as their BER has an error floor. Even though the BER for $\halfLifeT{0.004}$ has error floor, it manages to match the mutual information rate of other systems until $\tsym \simeq 0.024 ~s$. After that point, however, its capacity drops below the others because of the error floor on $\perr$ while all slower degradation cases converge to the same tail.

\subsubsection{Distance Analysis}

Degradation results in different behavior regarding channel capacity for various distances. Shortening of the time required to travel to the receiver allows the utilization of faster degradation, which in turn increases the channel capacity of the system. On the other hand, for longer distances the travel time is much delayed, therefore degradation may result in the destruction of all molecules and reduce the channel capacity of the system. 

\begin{figure}[t]
\centering
\includegraphics[width=0.55\columnwidth]{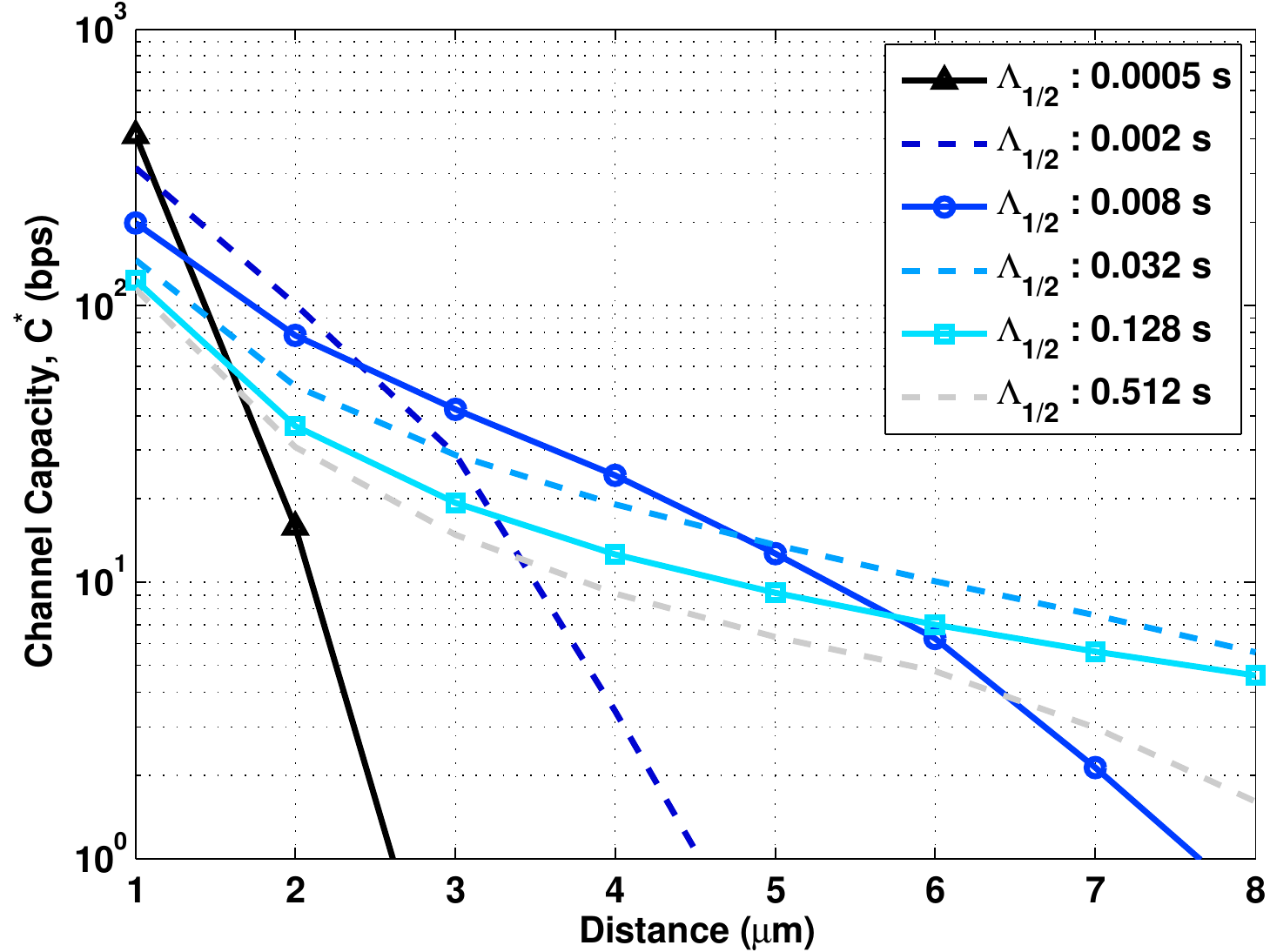}
\caption{The change in channel capacity with different distances at various degradation rates. For small distances, fast degradation rates significantly improve the capacity ($\times 3~ \text{for} ~d = 1~\mu m $ at $\halfLifeT{0.0005}$). However, as the communication distance gets larger fast degradation increases $\perr$ and significantly decreases $C^*$.}
\label{fig:Distance_vs_EDR}
\end{figure}

In Figure \ref{fig:Distance_vs_EDR}, we present $C^*$ with the optimal symbol duration for different distances. Note that $C^*$ is the peak points of the curves in Figure~\ref{fig:optimizeTs} while varying the distance. On smaller distances, best $C^*$ is achieved with faster degradation speeds (i.e., $\halfLifeT{0.0005}$), which clears the channel from the stray molecules better than slower ones. At longer distances, however, due to the travel time required, these fast degradation systems fail to achieve higher $C^*$ values.

\section{Conclusion}

In this paper, we provided a detailed analysis of non-enzymatic degradation in molecular communication via diffusion. In the analysis, we first analytically formulated the channel response function for an absorbing spherical receiver under molecular degradation. Second, utilizing this formulation, we clarified the signal-shaping aspect of degradation by investigating the effects of degradation on channel characteristics, namely, pulse peak time and pulse peak amplitude. Third, we examined the effects of degradation on continuous communication by evaluating system performance with respect to different communication parameters such as detection threshold, symbol duration, and distance of communication. The performance evaluation indicated that, by introducing the right level of degradation to the system, we can achieve notably better detection performance, reduced bit error rate, and increased channel capacity compared to systems without molecular degradation.

\section*{ACKNOWLEDGEMENT}

The work of A. C. Heren and T. Tugcu was supported in part by the State Planning Organization (DPT) of the Republic of Turkey under the project TAM with the project number 2007K120610, Bogazici University Research Fund (BAP) under grant number 7436, and by Scientific and Technical Research Council of Turkey (TUBITAK) under Grant number 112E011. The work of H. B. Yilmaz and C.-B. Chae was in part funded by the MSIP (Ministry of Science, ICT \& Future Planning), Korea, under the ``IT Consilience Creative Program" (NIPA-2014-H0201-14-1002) supervised by the NIPA (National IT Industry Promotion Agency) and by the Basic Science Research Program (2014R1A1A1002186) funded by the Ministry of Science, ICT and Future Planning (MSIP), Korea, through the National Research Foundation of Korea.

\bibliographystyle{IEEEtran}
\bibliography{IEEEabrv,bibliography}

\end{document}